\documentclass[twocolumn,prb,showpacs,preprintnumbers,amsmath,amssymb,superscriptaddress,bibliography]{revtex4-1}
\usepackage{graphicx}
\usepackage{setspace}

\usepackage{color}
\usepackage{ulem} 
\usepackage{makecell}
\graphicspath{{./figures/}}

\usepackage{amsmath,amsthm,amssymb}
\usepackage{mathrsfs}

\usepackage[
colorlinks=true, citecolor=blue, urlcolor=blue, linkcolor=blue,
setpagesize=false, bookmarks=false, breaklinks=true
]{hyperref}
\usepackage{lineno}

\newcommand{\hc}{\hat{c}}

\newcommand{\heta}{\hat{\eta}}

\newcommand{\hH}{\hat{H}}

\newcommand{\hn}{\hat{n}}

\newcommand{\hS}{\hat{S}}
\newcommand{\hP}{\hat{P}}

\newcommand{\bk}{{\bm T}}

\newcommand{\eqq}[1]{\begin{align} #1 \end{align}}

\begin{document}
\title{Many-body effects on high-harmonic generation in Hubbard ladders}

\author{Yuta Murakami}
\affiliation{Center for Emergent Matter Science, RIKEN, Wako, Saitama 351-0198, Japan}
\author{Thomas Hansen}
\affiliation{Department of Physics and Astronomy, Aarhus University, Ny Munkegade 120, DK-8000 Aarhus C, Denmark}
\author{Shintaro Takayoshi}
\affiliation{Department of Physics, Konan University, Kobe 658-8501, Japan}
\author{Lars Bojer Madsen}
\affiliation{Department of Physics and Astronomy, Aarhus University, Ny Munkegade 120, DK-8000 Aarhus C, Denmark}
\author{Philipp Werner}
\affiliation{Department of Physics, University of Fribourg, 1700 Fribourg, Switzerland}
\date{\today}

\begin{abstract} 
We show how many-body effects associated with background spin dynamics control the high-harmonic generation (HHG) in Mott insulators by analyzing the two-leg ladder Hubbard model.
Spin dynamics activated by the interchain hopping $t_y$ drastically modifies the HHG features. 
When two chains are decoupled ($t_y=0$), HHG originates from the dynamics of coherent doublon-holon pairs because of spin-charge separation.
With increasing $t_y$, the doublon-holon pairs lose their coherence due to their interchain hopping and resultant spin-strings.
Furthermore, the HHG signal from spin-polarons -- charges dressed by spin clouds -- leads to an additional plateau in the HHG spectrum.
For large $t_y$, we identify unconventional HHG processes involving {\it three} elementary excitations -- two polarons and one magnon.
Our results demonstrate the nontrivial nature of HHG in strongly correlated systems, and its qualitative differences to conventional semiconductors.

\end{abstract}

\maketitle
{\it Introduction--}
High-harmonic generation (HHG) is a fundamental and technologically important optical phenomenon originating from strong light-matter interaction~\cite{Corkum2007,Krausz2009RMP,Krausz2014}. In the past decade, HHG research has expanded dramatically from gases~\cite{Ferray_1988,Corkum1993PRL,Lewenstein1994} to solids~\cite{Ghimire2011NatPhys,Kruchinin2018RMP,Ghimire2019,Goulielmakis2022}, opening up new possibilities for optical and spectroscopic applications~\cite{Borsch2023}.
To date, most HHG research in solids has focused on semiconductors and semimetals~\cite{Ghimire2011NatPhys,Schubert2014,Vampa2014PRL,Luu2015,Vampa2015Nature,Langer2016Nature,Hohenleutner2015Nature,Ndabashimiye2016,Otobe2016,Tancogne-Dejean2017,Ikemachi2017,Liu2017,You2017,Kaneshima2018,Sekiguchi2022PRB,Giorgianni2016,Yoshikawa2017Science,Hafez2018,Silva2019,Chacon2020PRB,Matsunaga2020PRL,Schmid2021,Baykusheva2021}, where HHG is well described by the kinematics of independent electrons. Recently, the investigations have been further extended to correlated materials, where the independent-electron picture does not directly apply~\cite{Silva2018NatPhoton,Murakami2018PRL,Tancogne-Dejean2018,Freudenstein2022}.
Although previous studies have revealed some intriguing many-body effects, such as HHG from many-body states~\cite{Murakami2018PRL,Ishihara2020} and changes in the charge trajectory during the HHG process~\cite{Freudenstein2022}, 
 the broader implications of correlations on HHG are still not fully understood.
 
Strongly correlated systems (SCSs) provide a rich playground for studying many-body effects on HHG~\cite{Silva2018NatPhoton,Murakami2018PRL,Tancogne-Dejean2018,Ishihara2020,Murakami2021PRB,Orthodoxou2021,Udono2022PRB,Bionta2021PRR,Shao2022PRL,Hansen2022PRA,Uchida2022PRL,Murakami2022PRL,Hansen2022,Granas2022PRR,Alcala2022,nakano2023,Ono2024}. 
In SCSs, various degrees of freedom such as charge, spin, and orbital, are intertwined and the conventional band picture is not applicable~\cite{Dagotto1994RMP,Tokura_RMP,tdMott_revidw}.
Instead, the system can host intriguing excitation structures, which can lead to nontrivial HHG features.
Theoretically, SCSs are often represented by the Hubbard model, where strong local interactions lead to a Mott insulating state at half-filling.
Previous works revealed that HHG in Mott insulators can be associated with the coherent dynamics of many-body elementary excitations called doublons (doubly occupied sites) and holons (empty  sites)~\cite{Murakami2018PRL,Murakami2021PRB}. Furthermore, it was pointed out that strong spin-charge coupling can lead to an intriguing temperature dependence of HHG~\cite{Uchida2022PRL,Murakami2022PRL}.
However, only a few aspects of the spin-charge coupling have been explored, and potentially relevant effects, 
such as the formation of spin-strings~\cite{Ji2021PRX,Bohrdt2020PRB}, the emergence of new elementary excitations, or the scattering with magnons, have not been discussed.
A fundamental understanding of the role of  spin-charge coupling on HHG remains elusive.

Here, by analyzing the two-leg Hubbard model, we provide a comprehensive analysis of many-body effects on HHG originating from background spin dynamics in Mott insulators.
Due to its geometric structure, the two-leg system allows us to discuss physics beyond the previously studied one-dimensional-chain~\cite{Silva2018NatPhoton,Murakami2021PRB,Hansen2022PRA} and infinite-dimensional~\cite{Murakami2018PRL,Murakami2022PRL} limits.
Thus, the many-body effects revealed here should play an important role also in generic two- and three-dimensional systems, and be essential for the general understanding of HHG in SCSs.
  
\begin{figure}[t]
  \centering
    \hspace{-0.cm}
    \vspace{0.0cm}
\includegraphics[width=80mm]{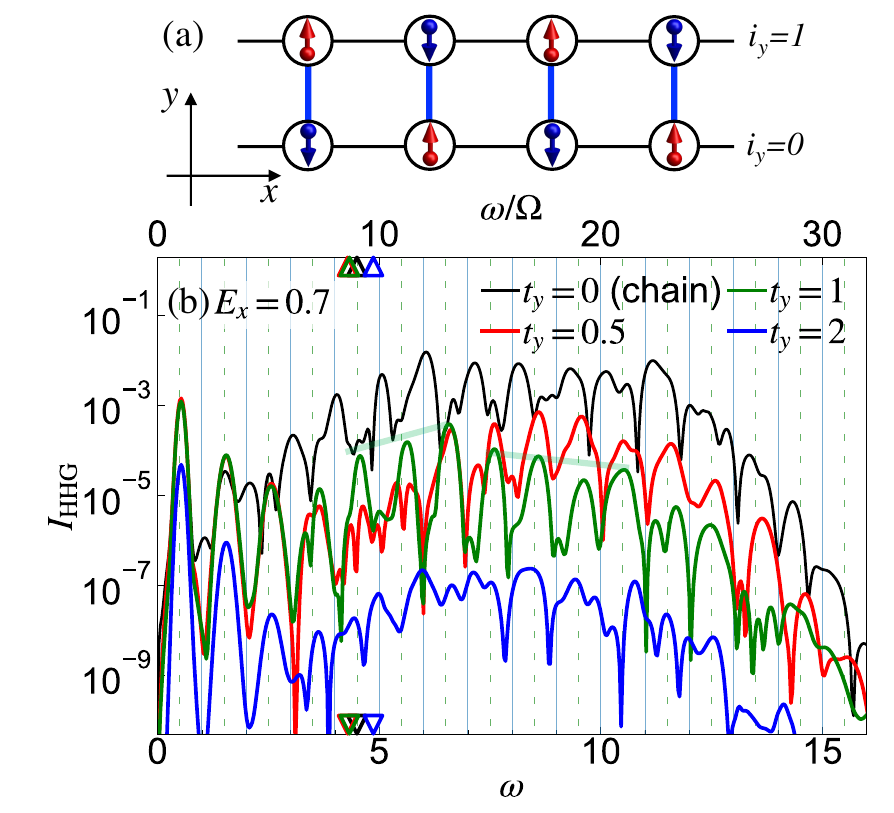} 
  \caption{(a) Schematic picture of the half-filled two-leg Hubbard model. (b) 
  HHG spectra for the half-filled two-leg Hubbard model for different $t_y$, see Eq.~\eqref{eq:Hubbard_2leg}.
    We use $t_x=1$ and $U=8$, and set  $\Omega=0.5,E_x=0.7, t_0=60$, and $\sigma_0=15$ for the electric field.
    The triangle markers indicate the optical gaps for each $t_y$, and the green transparent lines indicate the two plateau-like structures for $t_y=1$. }
  \label{fig:fig1}
\end{figure}

{\it Formulation--}
We focus on the half-filled Hubbard model defined on the two-leg ladder~\cite{Ladder_review,Noack1994PRL}
\eqq{
\hH(t)=&-t_x \sum_{i_x,i_y,\sigma} \big(e^{iA_x(t)}\hc^\dagger_{i_x+1 i_y \sigma} \hc_{i_xi_y\sigma} + h.c.\big) \nonumber \\
& - t_y \sum_{i_x,\sigma} \big(\hc^\dagger_{i_x0\sigma} \hc_{i_x1\sigma} + \text{h.c.}\big) +  \hH_U, \label{eq:Hubbard_2leg}
} 
see Fig.~\ref{fig:fig1}(a).
Here $\hc^\dagger_{i_xi_y\sigma}$ is the creation operator for an electron with spin $\sigma$, $i_x$ $(i_y)$ is the site index in the $x$ ($y$) direction, and $\hn_{i_xi_y\sigma}
=\hc^\dagger_{i_xi_y\sigma}\hc_{i_xi_y\sigma}$. $t_x$ and $t_y$ are the hopping parameters and $ \hH_U\equiv U \sum_{i_x,i_y} \hn_{i_xi_y\uparrow}\hn_{i_xi_y\downarrow}$ represents the on-site Coulomb interaction term.
The electric field is applied along $x$, and its effect is described by the Peierls phase with the vector potential $A_x(t)$. 
The corresponding electric field is $E_x(t) = -\partial_t A_x(t)$, and we use a Gaussian pulse $A_x(t)= \frac{E_{x}}{\Omega} F_G(t-t_0,\sigma_0) \sin(\Omega(t-t_0))$ with $F_G(t,\sigma)=\exp[-\frac{t^2}{2\sigma^2}]$.
We set the electronic charge $q$, the bond length, and $\hbar$ to unity. 
In the following, we use $t_x$ as the unit of energy and choose $U=8$, where the system is in the Mott insulating phase.
We use $\Omega=0.5,t_0=60$, and $\sigma_0=15$ for the excitation.
This parameter set is motivated by two-leg cuprates such as SrCu$_2$O$_3$~\cite{Muller1998PRB,Kuroki2020} excited by a mid-infrared laser,
for which our energy unit roughly corresponds to $0.5$~eV. 
We analyze the dynamics from the ground state using the infinite time-evolving block decimation (iTEBD)~\cite{Vidal2007PRL}.
The HHG spectrum is evaluated as $I_{\rm HHG}(\omega)=|\omega J_x(\omega)|^2$, where 
$J_x(\omega)$ is the Fourier component pf the current $J_x(t)$ along $x$, see Supplemental Material (SM) for details.
Due to the inversion symmetry, odd-harmonic peaks at $\omega=(2n+1)\Omega$ with $n\in\mathbb{N}$ are expected in fully time-periodic states~\cite{Cohen2019NatCom}.

{\it Results--}
Fig.~\ref{fig:fig1}(b) shows the HHG spectrum of the two-leg ladder for different values of $t_y$ with the field $E_x=0.7$.
For $t_y=0$ (two decoupled chains), peak structures at $\omega =(2n+1)\Omega$ are not clear and signals at non-odd-integer harmonics appears, see also SM for different $E_x$.
This indicates a remaining coherence between different Floquet states~\cite{Lange2024PRA}, so that the system is not yet fully time-periodic under the field.
We attribute this to the relatively short pump pulse and the long coherence time of the doublon-holon (DH) pairs due to the spin-charge separation~\cite{Murakami2021PRB}, which make the HHG spectrum sensitive to the carrier-envelope phase~\cite{Schmid2021}.
Meanwhile, with increasing $t_y$, HHG peaks become more prominent in general and develop around $(2n+1)\Omega$ (dashed lines), 
see the spectra for $t_y=0.5$ and $t_y=1$.
Additionally, signals just above the gap evolve non-monotonically as a function of $t_y$.  
For $t_y=1$, which is relevant for cuprates~\cite{Kuroki2020}, a new hump structure around $\omega=6.5$ develops, which results in two plateau-like structures above the gap ($4.5\lesssim \omega \lesssim 6.5$ and $6.5\lesssim \omega \lesssim 10.5$). The hump position increases with increasing field strength (see SM), indicating that this feature does not originate from a specific excitation mode.
When the interchain hopping $t_y$ is increased further, HHG peaks become less pronounced and non-odd-integer harmonics appear, as in $t_y=0$, suggesting a long coherence time of the excited charge carriers in this regime.

 \begin{figure}[t]
  \centering
    \hspace{-0.cm}
    \vspace{0.0cm}
\includegraphics[width=93mm]{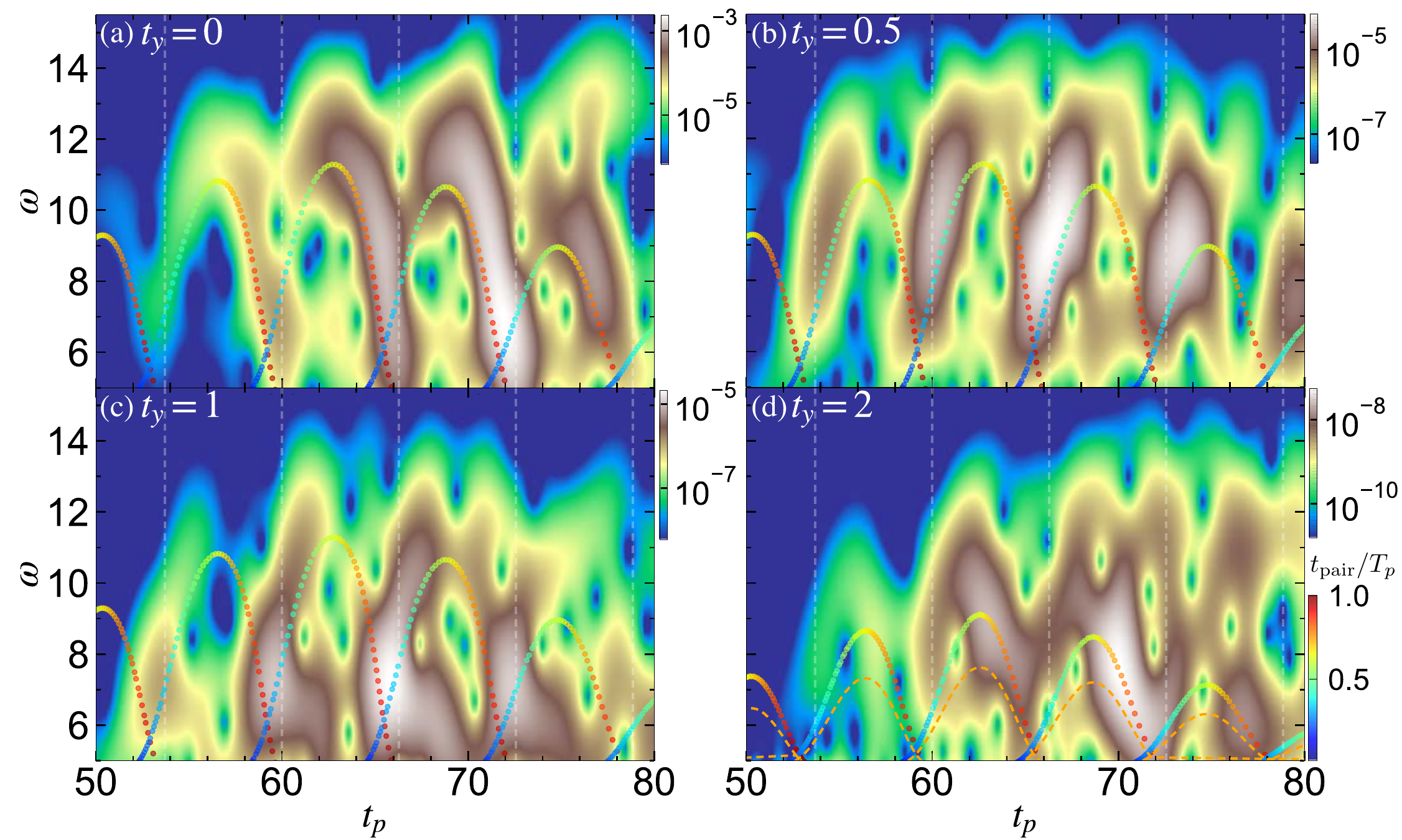} 
  \caption{Subcycle spectra corresponding to the data shown in Fig.~\ref{fig:fig1}(b), based on a
 Gaussian window with $\sigma_p=0.8$. The vertical dashed lines indicate the times with $A_x(t)=0$.
 The multi-colored dots and the orange dashed lines indicate the energy emission at time $t_p$ within the three-step model. 
 In panels (a-c) the dispersion of the DH pair is estimated from the chain system.
In (d), the dots are for Scenario 1 (hPolaron + ePolaron + magnon), while the orange dashed line is for Scenario 2 (hPolaron + spin-bag).
The color bar covers the time interval between creation and recombination of the corresponding pair in units of $T_p=\frac{2\pi}{\Omega}$.
}
  \label{fig:fig2}
\end{figure}

To clarify the effects of the interchain hopping, we conduct a subcycle analysis. Specifically, we apply the windowed Fourier transform 
$J_x(\omega,t_p)=\int dt e^{i\omega t} F_G(t-t_p,\sigma_p)J_x(t)$ and extract the temporal radiation spectrum around $t_p$ as $I(\omega,t_p)=|\omega J_x(\omega,t_p)|^2$.
The results are shown in Fig.~\ref{fig:fig2}, where the rainbow-colored dots in Figs.~\ref{fig:fig2}(a-c) represent the prediction from the semiclassical three-step model for the DH pair~\cite{Murakami2021PRB}, see SM.
Here, the DH dispersion for the chain ($t_y=0$) is used.  The dot color indicates the time between the creation and annihilation of the corresponding DH pair.
For $t_y=0$, the temporal change of the radiation intensity is well described by the three-step model and the radiation is mainly emitted by long-lived DH pairs. 
With increasing $t_y$, we first notice a reduction of the DH coherence, as manifested by the shift of the weight of $I(\omega,t_p )$ to the subcycle interval corresponding to short-lived DH pairs, see $t_y=0.5$. For $t_y=1$, the chain-like DH trajectory of the (short-lived) DH pair is still visible, but simultaneously there emerges a low-lying dispersive signal around $\omega \approx 6$. The reduction of the coherence with increasing $t_y$ is consistent with the development of the clear HHG peaks since it allows the system to quickly reach a time-periodic state.
The low-lying dispersive signal can be associated with the field-driven dynamics of certain elementary excitations,
and it explains the first hump structure in the HHG spectrum.
For $t_y=2$, the weight of $I(\omega,t_p)$ is shifted back to later in the cycle of the electric field, which suggests the existence of photo carriers with long coherence and is consistent with the observed features of the HHG spectrum.

Since the drastic changes in the qualitative features of HHG with interchain hopping are absent in conventional semiconductors (see SM), 
they are linked to the strongly-correlated nature of the system. 
In the following, we demonstrate that these changes manifest different aspects of the spin-charge coupling.

First, we reveal the intriguing mechanism of the reduction of the DH coherence.
In Ref.~\cite{Murakami2022PRL}, such a reduction due to the spin-charge coupling was already pointed out. 
That study, however, focused on the energetic coupling between spin and charge, which converts the kinetic energy of a doublon or holon into spin exchange energy via the disturbance of the spin background. 
Still, the effect of this coupling is expected to be small for $t_y=0.5$, where the spin-exchange energy is $J_{\text{ex},y}=0.125$, 
and there must be other mechanisms which control the coherence time. 
To clarify the origin of the observed dephasing, we evaluate the HHG using an effective model, which is obtained by the Schrieffer-Wolff transformation, i.e. an expansion in $t_{x,y}/U$~\cite{MacDonald1988,Murakami2021PRB,Murakami2022}.
The Hamiltonian reads 
\eqq{
\hH_{\rm eff}(t) = \hH_{\rm DH}(t) + \hH_{\rm spin} + \hH_U- E_x(t) \hat{P}_x(t), \label{eq:Heff}
}
where 
 \eqq{
&\hH_{\rm DH}(t) =   -t_y\sum_{i_x,\sigma} \hat{\bar{n}}_{i_x0\bar{\sigma}} (\hc^\dagger_{i_x0\sigma}\hc_{i_x1\sigma} + h.c.)  \hat{\bar{n}}_{i_x1\bar{\sigma}} \nonumber \\
& -t_y\sum_{i_x,\sigma} \hn_{i_x0\bar{\sigma}} (\hc^\dagger_{i_x0\sigma}\hc_{i_x1\sigma} + h.c.) \hn_{i_x1\bar{\sigma}} \nonumber \\
& -t_x\sum_{i_x,i_y,\sigma}  \hat{\bar{n}}_{i_x+1i_y\bar{\sigma}}  (e^{iA_x(t)}\hc^\dagger_{i_x+1i_y\sigma}\hc_{i_xi_y\sigma} + h.c.) \hat{\bar{n}}_{i_xi_y\bar{\sigma}}\nonumber \\
& -t_x\sum_{i_x,i_y,\sigma} \hn_{i_x+1i_y\bar{\sigma}}   (e^{iA_x(t)}\hc^\dagger_{i_x+1i_y\sigma}\hc_{i_xi_y\sigma} + h.c.)  \hn_{i_xi_y\bar{\sigma}} \nonumber 
 }
 represents the hopping of doulons and holons. $\bar{\sigma}$ indicates the opposite spin of $\sigma$ and  $ \hat{\bar{n}} = (1-\hn)$.
 $ \hH_{\rm spin}= J_{{\rm ex},x} \sum_{i_x,i_y} \hat{\bf  s}_{i_xi_y}\cdot \hat{\bf  s}_{i_xi_y} + J_{{\rm ex},y} \sum_{i_x} \hat{\bf  s}_{i_x0}\cdot \hat{\bf  s}_{i_x1}$ is the spin exchange term with $J_{{\rm ex},x} = \frac{4 t_x^2}{U}$ and $J_{{\rm ex},y} =  \frac{4 t_y^2}{U}$. $\hat{P}_x(t)$ creates or annihilates a DH pair and acts as an interband dipole moment, see SM for the expression. 
 This model is an extension of the $t$-$J$ model~\cite{Dagotto1994RMP} to a system with both doublons and holons, and allows us to separately evaluate the effects of different physical processes~\cite{Murakami2021PRB}.
We prepare the initial state $|\psi_\text{in}\rangle$ as the ground state of $\hH_{\rm eff}$, and then evolve the system from $|\psi_\text{in}\rangle$ with the following modified coefficients in $\hH_{\rm eff}(t)$ \eqref{eq:Heff}: 1) $t_y=J_{\rm ex,x}=J_{\rm ex,y}=0$,  2) $J_{\rm ex,x}=J_{\rm ex,y}=0$, 3) $t_y=0$, and 4) no change. Since $|\psi_\text{in}\rangle$ has no doublons and holons, it is essentially determined by the spin Hamiltonian $\hH_{\rm spin}$, and the four cases share $|\psi_\text{in}\rangle$ as the ground state. 
 The comparison of the results for 1)-4) provides insights into the role of different physical processes. 
Namely, the difference between 1) and 2) shows the effects of the DH dynamics along $y$ without energy cost from the spin exchange coupling.
Meanwhile, the difference between 1) and 3) reveals the effect of the energetic coupling originating from the spin mismatch between the chains (the effect discussed in Ref.~\cite{Murakami2022PRL}).

 \begin{figure}[t]
  \centering
    \hspace{-0.cm}
    \vspace{0.0cm}
\includegraphics[width=90mm]{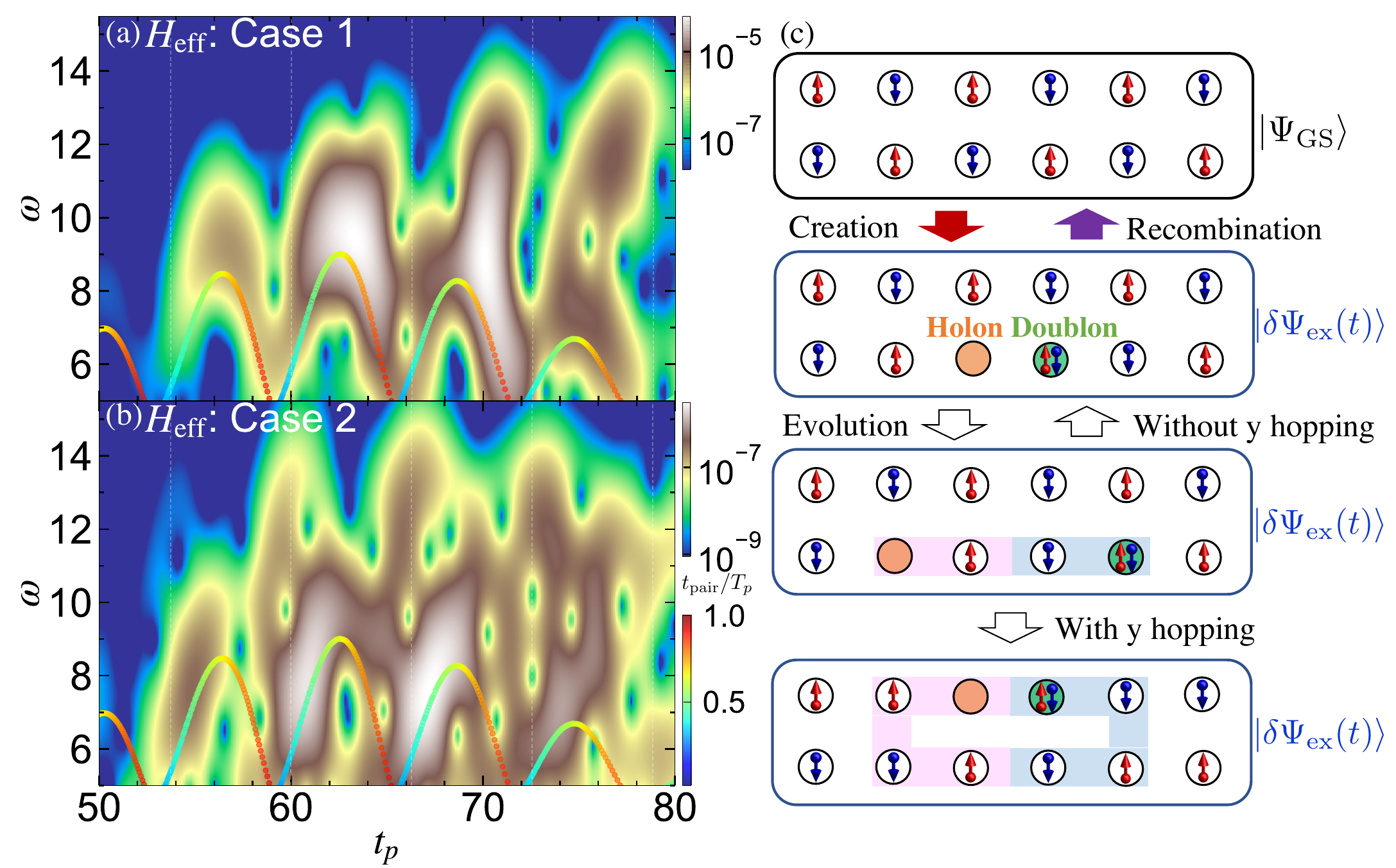} 
  \caption{(a,b) Subcycle spectra of the two-leg Mott insulator described by the effective model.
    We use $t_x=1$, $t_y=0.5$, $U=8$ for the system, $\Omega=0.5$, $E_x=0.7$, $t_0=60$, $\sigma_0=15$ for the field pulse, and $\sigma_p=0.8$ for the analysis.
    The rainbow-colored dots are the prediction from the three-step model of the DH pair, whose dispersion is $E_g(k_x) = U -4t_x\cos(k_x)$.
    (c) Schematic illustration of the effects of DH hopping along $y$ on the light emission. The shaded area indicates the trajectory of the doublon and holon, i.e., the formation of a string in the spin background, which prohibits the return to the ground state by DH recombination. }
  \label{fig:fig3}
\end{figure}

In Figs.~\ref{fig:fig3}(a-b), we show the resultant subcycle spectrum for Cases 1) and 2). The colored line shows the prediction from the three-step model with the DH dispersion of $U-4t_x\cos(k_x)$. 
In Case 1), where only the DH motion along $x$ is considered, the DH pairs exhibit a long coherence time. 
The DH hopping along $y$ (Case 2)) causes a significant shift of the weight to the early part of the cycle, implying a substantial reduction of DH pair coherence. 
Although a reduction in coherence is also observed in Case 3), for the present parameter set, the effect is much more prominent in Case 2), see SM.

The reduction of pair coherence in Case 2) can be attributed to the existence of spin-strings (disturbances of the spin configuration from the ground state)~\cite{Ji2021PRX, Bohrdt2020PRB}, see Fig.~\ref{fig:fig3}(c).
Here we express the time-evolving wave function as the sum of the ground state and the excited state, $|\Psi(t)\rangle = |\Psi_{\rm GS}\rangle + |\delta\Psi_{\rm ex}(t)\rangle$.
The main contribution to the current comes from $\langle \Psi_{\rm GS} |\hat{J}_x(t) |\delta \Psi_{\rm ex}(t) \rangle$, which involves the DH recombination.
$\hat{J}_x(t) \equiv - \frac{\delta \hH(t)}{\delta A_x(t)}$ is the current operator.
In Case 1), without DH hopping along $y$, the spin background remains close to that of the ground state after the DH pair has moved around.
Thus, when the DH pair returns to neighboring sites, it contributes to $\langle \Psi_{\rm GS} |\hat{J}_x(t) |\delta \Psi_{\rm ex}(t) \rangle$.
In contrast, in Case 2), if hopping along $y$ happens during the dynamics,
spin-strings remain even after the DH pair returns to neighboring sites, see the bottom panel of Fig.~\ref{fig:fig3}(c).
Due to the spin-strings, the pair cannot recombine, and does {\it not} contribute to $\langle \Psi_{\rm GS} |\hat{J}_x(t) |\delta \Psi_{\rm ex}(t) \rangle$, which results in a reduction of pair coherence.
Importantly, this process is active even in the limit of $J_\text{ex}=0$, where there is no energetic coupling between spin and charge.
It is absent in semiconductors, due to the lack of spin background dynamics, see SM.

 \begin{figure}[t]
  \centering
    \hspace{-0.cm}
    \vspace{0.0cm}
\includegraphics[width=90mm]{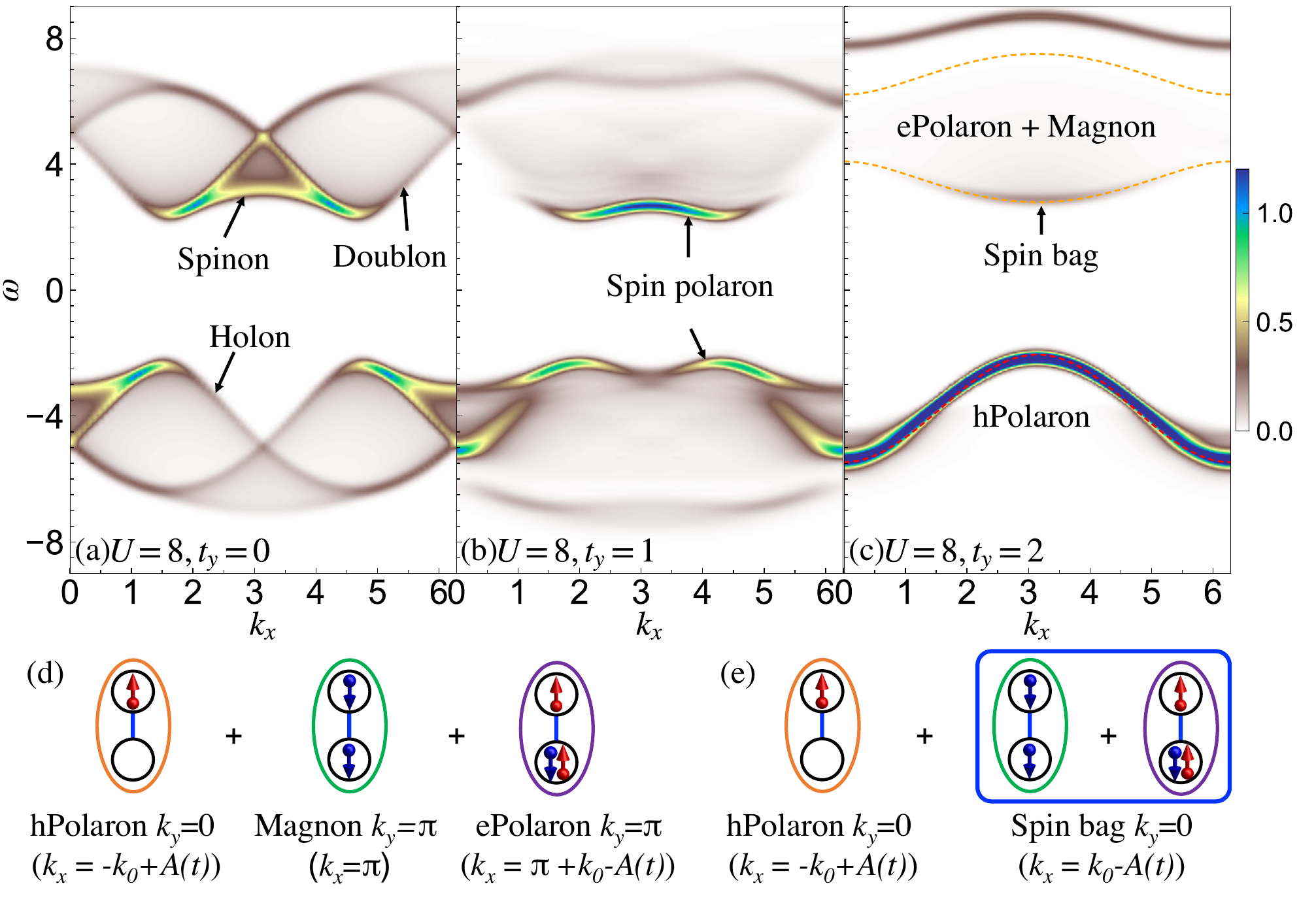} 
  \caption{(a-c) Single-particle spectra $A(k_x,k_y=0,\omega)$ of the half-filled two-leg Hubbard model calculated with DMRG.
  We set $t_x=1$ and $U=8$ and the number of sites along $x$ is $80$. 
  The dashed lines in (c) indicate the dispersion of the hPolaron and the range of the continuum consisting of an ePolaron and a magnon, 
  as predicted by the strong-rung perturbation theory. (d,e) Two different scenarios of the HHG process in the strong-rung regime.
  }
  \label{fig:fig4}
\end{figure}

Next, we discuss the effects of emergent elementary excitations on HHG. 
To obtain hints on relevant elementary excitations for $t_y=1,2$, we analyze the single-particle spectra $A(k_x,k_y,\omega)$ in momentum space,
which captures the states accessible by removing (adding) an electron from (to) the system~\cite{Feiguin2019PRB}.
In Fig.~\ref{fig:fig4}, we show $A(k_x,k_y,\omega)$ for $k_y=0$, obtained with the time-dependent density-matrix renormalization group (DMRG)~\cite{White2004PRL,Haegeman2011PRL}.
In the present system, $k_y$ is either $0$ or $\pi$, and $A(k_x,k_y,\omega)=A(k_x+\pi,k_y+\pi,-\omega)$.
For $t_y=0$, dispersive signals corresponding to doublons, holons and spinons are observed~\cite{Essler2005,Bohrdt2018}.
For $t_y=1$, a low-lying dispersive band ($2.5 \lesssim |\omega|\lesssim 3.5$) corresponding to the formation of spin-polarons
(charges dressed by spin clouds) emerges~\cite{Feiguin2019PRB}.
For $t_y=2$, we see a coherent band for $\omega<0$ and a continuum around $2.5\lesssim \omega\lesssim 7.5$.
For $t_y=2$, perturbation theory in the limit of strong-rung coupling works well, see SM. 
The band for $\omega<0$ corresponds to a polaron with a positive charge $-q$ (hPolaron) and $k_y=0$.
Meanwhile, the continuum consists of a polaron with a negative charge $q$ (ePolaron) with $k_y=\pi$ and a magnon with $k_y=\pi$.
The signal at the bottom of the continuum has high intensity. This signal represents a weakly-bound state of an ePolaron and a magnon, a composite particle called a spin-bag~\cite{Feiguin2019PRB}.

We now reveal the role of these elementary excitations.
For $t_y=1$, the recombination of two polarons can yield light in the range of $5 \lesssim \omega_{\rm emit}\lesssim 7$,
which matches the range of the first plateau observed in HHG. 
Thus, the plateau can be attributed to the coherent dynamics of two spin-polaron pairs.
In other words, the electron dynamics hosts both aspects of DH dynamics and polaron dynamics at the same time due to correlation effects, and HHG simultaneously captures them. 
For $t_y=2$, the following two scenarios based on the three-step picture are conceivable.
Since we excite the system with a homogeneous field, the total momentum of the relevant elementary excitations must remain zero.
Scenario 1 involves three elementary excitations, a hPolaron with $k_y=0$, an ePolaron with $k_y=\pi$, and a magnon $k_y=\pi$, see Fig.~\ref{fig:fig4}(d).
This situation is very different from the conventional scenario, which involves two elementary excitations~\cite{Vampa2015PRB}.
Since a magnon carries no charge, only the polarons move around under the electric field to acquire kinetic energy.
Still, the magnon balances the total momentum in the tunneling and recombination processes, and yields an energy shift corresponding to its creation energy.
Scenario 2, involves two elementary excitations, i.e. one hPolaron and one spin bag, see Fig.~\ref{fig:fig4}(e). 
To check the validity of these scenarios, in Fig.~\ref{fig:fig2}(d), we compare the subcycle spectrum and the prediction from the semiclassical theory for the two scenarios.
The comparison shows that Scenario 1 provides a better match with the numerical results, see also SM. 
Scenario 1 consistently explains that the recovery of the coherence of the signal originates from the emergence of coherent polarons expected in the strong-rung regime.
Still, we note that in $A(k_x,k_y,\omega)$, the $\omega>0$ part looks incoherent and information on the ePolaron is hidden.
This result shows that the relation between HHG and the single-particle spectrum is not straightforward, in contrast to semiconductors.~\cite{Murakami2021PRB}
In general, HHG provides direct access to the kinematics of elementary excitations with nonzero charge.

{\it Discussion---}
Our study of the two-leg Hubbard model revealed various many-body effects on HHG associated with the spin background dynamics.
Experimentally, these characteristic HHG features could be systematically studied in two-leg systems \cite{Ladder_review,Tokura_RMP,Abbamonte2004} such as SrCu$_2$O$_3$ by tuning the ratio $t_y/t_x$ through chemical and physical pressure~\cite{Kuroki2020}.
Larger modifications of the hopping parameters can be achieved with nonlinear phononics~\cite{Forst2011}, i.e., resonant optical excitations of the relevant phonons. This technique can produce lattice displacements beyond the material's breaking point under static pressure.
Detailed information on the subcycle dynamics may be extracted using ultrafast techniques such as the time-domain observation of electric fields~\cite{Park2018optica,Cho2019,Madsen2023PRB}, attosecond transient-absorption spectroscopy~\cite{Wu2016ATAS,Madsen2015PRA}, and multidimensional spectroscopy~\cite{Valmispild2024}.

The many-body effects discussed here should not be limited to two-leg systems.
Considering the similarity in the geometry and the excitation structures~\cite{Ji2021PRX,Feiguin2016PRB,Bohrdt2020PRB}, the effects of spin-strings as well as spin-polarons on HHG 
should be highly relevant even in two-dimensional systems. 
We also expect that HHG involving multiple elementary excitations is common in correlated systems.
Our results serve as an essential step toward a microscopic understanding of nontrivial HHG features, as reported in various strongly correlated materials~\cite{Alcala2022,nakano2023}.

\begin{acknowledgments}
We thank T. Ikeda, K. Sugimoto, T. Kaneko and K. Shinjo for the fruitful discussion.
This work is supported by Grant-in-Aid for Scientific Research from JSPS, KAKENHI Grant Nos. JP21H05017 (Y.M.), JP24H00191(Y.M and S.T), JP23K22418 (S.T.), JP24K06891 (S.T.), JST CREST Grant No. JPMJCR1901 (Y.M.), the RIKEN TRIP initiative RIKEN Quantum (Y.M.), the independent Research Fund Denmark Grant No. 1026-00040B (T.H., L.B.M.), and SNSF Grant  No. 200021-196966 (P.W.). The iTEBD calculations have been implemented using the open-source library ITensor~\cite{ITensor_main,ITensor_code}.
\end{acknowledgments}

\appendix
\section{Details of the numerical simulations}
We focus on the two-leg Hubbard model
 \eqq{
\hH(t)&=-t_x \sum_{{\bm i},\sigma} (e^{iA_x(t)}\hc^\dagger_{{\bm i} + {\bm e}_x \sigma} \hc_{{\bm i}\sigma} + h.c.)\label{eq:Hubbard_2leg}  \\
& - t_y \sum_{i_x,\sigma} (\hc^\dagger_{i_x0\sigma} \hc_{i_x1\sigma} + h.c.) + U \sum_{\bm{i}} \hn_{\bm i \uparrow}\hn_{\bm i\downarrow}-\mu\sum_{\bm i} \hn_{\bm i} , \nonumber 
} 
where we explicitly added the chemical potential $\mu$.
The notation is the same as in the main text. In particular, ${\bm i}=(i_x,i_y)$ is a unified site index and ${\bm e}_x=(1,0)$ is the unit vector along $x$.
The first two terms define the hopping term $\hH_{\rm kin}$, while $\hH_U\equiv U \sum_{\bm{i}} \hn_{\bm i \uparrow}\hn_{\bm i\downarrow}$ represents the Hubbard interaction.
We analyze the system numerically using the following methods based on the matrix product state (MPS) representation of the wave function. 

\subsection{Infinite time-evolving block decimation (iTEBD)}
In iTEBD, the translational invariance of the MPS is exploited, which allows us to treat the thermodynamic limit directly~\cite{Vidal2007PRL}.
The imaginary-time and real-time evolution of the MPS is implemented using the Trotter-Suzuki decomposition of the target Hamiltonian.
In particular, the ground state is obtained with imaginary time evolution.

 \begin{figure}[t]
  \centering
    \hspace{-0.cm}
    \vspace{0.0cm}
\includegraphics[width=85mm]{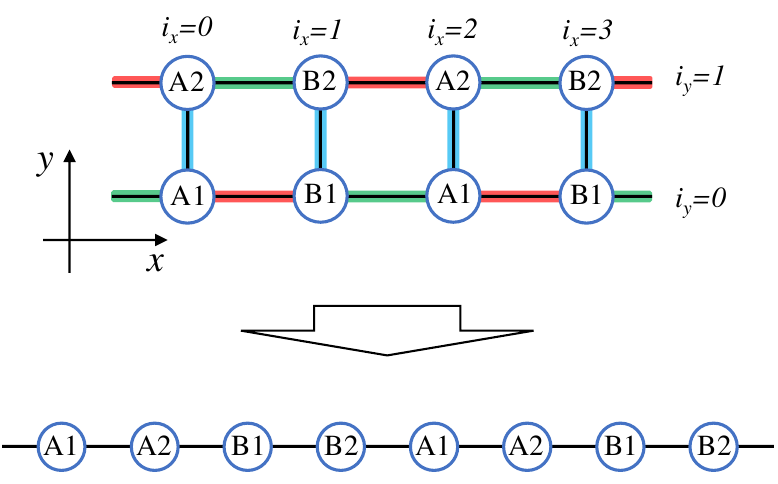} 
  \caption{Illustration of the two-leg Hubbard model. The upper part shows the original configuration, while the lower part shows the configuration 
  transformed into a 1D chain. The different sets of bonds, on which $\hH_A$, $\hH_B$ and $\hH_C$ of Eq.~\eqref{eq:H_ABC} are defined, are indicated with the different colors in the upper panel.
}
  \label{fig:S_2leg}
\end{figure}

In the actual implementation of iTEBD for the two-leg ladder system \eqref{eq:Hubbard_2leg},
we regard it as a one-dimensional (1D) single-chain system with next-nearest neighbor terms in the Hamiltonian, see Fig.~\ref{fig:S_2leg}.
Such terms can be treated with the swap operator~\cite{Stoudenmire2010}.
In practice, we separate  Eq.~\eqref{eq:Hubbard_2leg} into three terms connecting different bonds (see Fig.~\ref{fig:S_2leg}) as 
\eqq{
\hH_A =& -t_x \sum_{i_x+i_y={\rm even},\sigma}  (e^{iA_x(t)}\hc^\dagger_{{\bm i} + {\bm e}_x \sigma} \hc_{{\bm i}  \sigma} + \text{h.c.}), \nonumber \\
\hH_B =&  - t_y \sum_{i_x,\sigma} (\hc^\dagger_{i_x0\sigma} \hc_{i_x1\sigma} + \text{h.c.}) \nonumber \\
 &+ U \sum_{\bm{i}} \hn_{\bm{i}\uparrow}\hn_{\bm{i}\downarrow}-\mu\sum_{\bm{i}} \hn_{\bm{i}},  \label{eq:H_ABC}  \\
\hH_C =& -t_x \sum_{i_x+i_y={\rm odd},\sigma}  (e^{iA_x(t)}\hc^\dagger_{{\bm i} + {\bm e}_x \sigma} \hc_{{\bm i} \sigma} + \text{h.c.}). \nonumber 
}
Note that there is a degree of freedom to partially move the local Coulomb interaction or the chemical potential term from $\hH_B$ to $\hH_A$ and $\hH_C$.
We apply the fourth-order Trotter-Suzuki decomposition for these three non-commuting terms~\cite{Ostmeyer2023,Barthel2020} using the swap operators.
For the real-time evolution, we additionally use fourth-order commutator-free matrix exponentials~\cite{Alvermann2011}.
Our code is based on the open library ITensor~\cite{ITensor_main,ITensor_code}.

We evaluate the HHG spectrum from the current induced by the electric field along $x$.
The current operator along $x$ is defined as 
\eqq{
\hat{J}_x(t) = \frac{it_x}{N} \sum_{\bm{i},\sigma} \Big[e^{iA_x(t)}  \hc^\dagger_{\bm{i}+{\bm e}_x\sigma}\hc_{\bm{i}\sigma} - e^{-iA_x(t)} \hc^\dagger_{\bm{i}\sigma}\hc_{\bm{i}+{\bm e}_x\sigma}\Big], \label{eq:current}
}
which is normalized by the number of sites $N(=N_x\times 2)$.  The HHG spectrum is calculated as
\eqq{
I_{\rm HHG}(\omega) = |\omega J_x(\omega)|^2.
}
Here $J_x(\omega)$ is the Fourier component of $J_x(t)\equiv \langle \hat{J}_x(t) \rangle$.
Since the numerical simulation is limited to a finite time-range $[0,t_{\rm max}]$, it is useful to introduce a Gaussian window 
$F_{\rm G}(t-t_0,\sigma) = \exp\bigl(-\frac{(t-t_0)^2}{2\sigma^2}\bigl)$ which is wide enough compared to the pulse width but shorter than $t_{\rm max}-t_0$ and $t_0$.
Note that $t_0$ is the center of the electric field pulse. 
In this paper, we use $\sigma=20$ for the excitation with $\Omega=0.5,t_0=60$ and $\sigma_0=15$, but we found that the shape of the HHG spectrum does not sensitively depend on the choice of $\sigma$.

For the single chain ($t_y=0$), we prepare the ground state by the imaginary time evolution with the cut-off dimension $D=1000$ and we set $D=3000$ and the time step $dt=0.04$ for the real-time evolution, unless otherwise mentioned.
To reduce the cut-off dimension, we add a tiny staggered magnetic field  ($h_z=0.001$) for the single chain simulation. 
For the two-leg ladders with nonzero $t_y$, we use $D=2000$ for the preparation of the ground state, and $D=3000$ and $dt=0.04$ for the real-time evolution, unless otherwise mentioned. These conditions show reasonable convergence of the HHG features.

\subsection{Time-dependent DMRG} \label{sec:DMRG}
In the time-dependent density matrix renormalization group (DMRG) approach, the ground state is first prepared by the usual DMRG~\cite{ITensor_main}. 
The subsequent time evolution is implemented using the Trotter-Suzuki decomposition as in the TEBD method~\cite{White2004PRL}
or using the time-dependent variational principle (TDVP)~\cite{Haegeman2011PRL,Haegeman2016PRB}.
We numerically confirmed that both methods lead to essentially the same results in our case.
For the time-dependent DMRG, we also use ITensor~\cite{ITensor_main,ITensor_code}.

Time-dependent DMRG is used to evaluate the Green's functions $G^R_{\bm{ij}\sigma}(t) = -i\theta(t)\langle [\hc_{{\bm i}\sigma}(t),\hc^\dagger_{{\bm j}\sigma}(0)]_+\rangle$.
Here, ${\bm i}$ and ${\bm j}$ denote the site indices and $[,]_+$ the anti-commutator. In practice, we prepare the ground state $|\psi_{\rm GS}\rangle$ for a system with open boundary condition and apply $\hc^\dagger_{{\bm j}\sigma}$ at the center of the system as $\hc^\dagger_{{\bm j}\sigma}|\psi_{\rm GS}\rangle$. 
We simulate the time evolution of the state up to time $t$ and evaluate  $\langle \hc_{{\bm i}\sigma}(t) \hc^\dagger_{{\bm j}\sigma}(0)\rangle$. 

The obtained $G^R_{\bm{ij}\sigma}(t)$ is used to evaluate the single-particle spectrum $A(\bk,\omega) = -\frac{1}{\pi} {\rm Im} G^R(\bk,\omega)$.
Here $G^R(\bk,\omega)$ is the Fourier transform of $G^R_{\bm{ij}\sigma}(t)$ with respect to space and time.

\section{Effective model from the Schrieffer-Wolff transformation} \label{sec:SW_model}
In this section, we apply the Schrieffer-Wolff (SW) transformation assuming $t_y,t_x\ll U$ to derive the effective model Eq.~(2) in the main text.
This model was previously used for the analysis of HHG in the 1D single-chain Hubbard model~\cite{Murakami2021PRB}
and allows us to separate different physical processes.

The (time-dependent) SW transformation can be expressed as 
 \eqq{
 \hat{H}_{\rm SW}(t) = e^{i\hS(t)} \hH e^{-i\hS(t)} + i(\partial_t e^{i\hS(t)}) e^{-i\hS(t)}. \label{eq:SW}
 }
$\hS(t)$ is expanded in terms of $t_{xy}/U$ as 
 \eqq{
 \hS = \hS^{(1)} + \hS^{(2)} + \hS^{(3)} + \cdots.
 }
 Here, $t_{xy}$ indicates the typical energy scale of $t_x$ and $t_y$.
$\hS^{(i)}$ denotes the terms of order $(t_{xy}/U)^i$, which are determined such that at each order $i$ there is no doublon-holon (DH) creation/annihilation in $e^{i\hS(t)} \hH e^{-i\hS(t)}$. 
Effective models can be obtained by truncating this Hamiltonian at a given order.

 One can systematically construct $\hS^{(i)}$ as explained in Refs.~\cite{MacDonald1988,Murakami2021PRB,Murakami2022}. To this end, we separate the hopping term of the Hubbard model  ($\hH_{\rm kin}(t)$), 
 into four terms describing the dynamics of the doublons and holons: 
 \eqq{
 \hH_{\rm kin, H}(t) =& -t_x\sum_{\bm{i},\sigma} \hat{\bar{n}}_{\bm{i}\bar{\sigma}} (e^{iA_x(t)}\hc^\dagger_{\bm{i}+{\bm e}_x\sigma}\hc_{\bm{i}\sigma} + h.c.)  \hat{\bar{n}}_{\bm{i}+{\bm e}_x\bar{\sigma}} \nonumber \\&  -t_y\sum_{i_x,\sigma} \hat{\bar{n}}_{i_x0\bar{\sigma}} (\hc^\dagger_{i_x0\sigma}\hc_{i_x1\sigma} + h.c.)  \hat{\bar{n}}_{i_x1\bar{\sigma}},\\ 
 \hH_{\rm kin, D}(t) =& -t_x\sum_{\bm{i},\sigma} \hn_{\bm{i}\bar{\sigma}} ( e^{iA_x(t)}\hc^\dagger_{\bm{i}+{\bm e}_x\sigma}\hc_{\bm{i}\sigma} + h.c.)  \hat{n}_{\bm{i}+{\bm e}_x\bar{\sigma}} \nonumber 
 \\ & -t_y\sum_{i_x,\sigma} \hn_{i_x0\bar{\sigma}} (\hc^\dagger_{i_x0\sigma}\hc_{i_x1\sigma} + h.c.)  \hn_{i_x1\bar{\sigma}}, \\
 \hH_{\rm kin, + }(t)= & -t_x\sum_{\bm{i},\sigma} [\hn_{\bm{i}\bar{\sigma}} e^{-iA_x(t)}\hc^\dagger_{\bm{i}\sigma}\hc_{\bm{i}+{\bm e}_x\sigma}  \hat{\bar{n}}_{\bm{i}+{\bm e}_x\bar{\sigma}}   \nonumber \\
 &\;\;\;\;\;\;\;\;\;\;\;+ \hn_{\bm{i}+{\bm e}_x\bar{\sigma}}  e^{iA_x(t)}\hc^\dagger_{\bm{i}+\bm{e}_x\sigma} \hc_{\bm{i}\sigma}  \hat{\bar{n}}_{\bm{i}\bar{\sigma}}] \nonumber \\ 
 & -t_y\sum_{i_x,\sigma} [\hn_{i_x0\bar{\sigma}} \hc^\dagger_{i_x0\sigma}\hc_{i_x1\sigma}  \hat{\bar{n}}_{i_x1\bar{\sigma}}   \nonumber \\
 &\;\;\;\;\;\;\;\;\;\;\; + \hn_{i_x1\bar{\sigma}}  \hc^\dagger_{i_x1\sigma}\hc_{i_x0\sigma}  \hat{\bar{n}}_{i_x0\bar{\sigma}}], \\
 \hH_{\rm kin,-}(t) =& \hH_{\rm kin,+}(t)^\dagger.
 }
 Here, $\bar{\sigma}$ indicates the spin opposite to $\sigma$ and  $ \hat{\bar{n}} = (1-\hn)$.
 $ \hH_{\rm kin, H}(t)$ ($ \hH_{\rm kin, D}(t)$) represents the hopping of a holon (doublon) which does not change the number of doublons or holons.
 On the other hand, $\hH_{\rm kin, + }(t)$ ($ \hH_{\rm kin,-}(t)$) represents a motion which creates (annihilates) a DH pair.

First, we consider the behavior of $e^{i\hS(t)} \hH(t) e^{-i\hS(t)}(\equiv \hH'(t))$.
 Since $ e^{i\hS} \hH e^{-i\hS} = \hH + i [\hS,\hH] - \frac{1}{2} [\hS,[\hS,\hH]] + \cdots$,
 the term at a given order is expressed as 
 \eqq{
  \hH'^{(0)} &= \hH_U, \\
  \hH'^{(1)}(t) & =   \hH_x(t) +   \hH_{y} + i[\hS^{(1)}(t),\hH_U] ,\\
    \hH'^{(2)}(t) & =  i [\hS^{(1)}(t),\hH_{\rm kin}(t)] + i[\hS^{(2)}(t),\hH_U] \nonumber  \\
            & \;\;\; -\frac{1}{2} [\hS^{(1)}(t),[\hS^{(1)}(t),\hH_U]].
 }

For the leading order, we can set
 \eqq{
 \hS^{(1)}(t) = -\frac{i}{U}[\hH_{{\rm kin},+}(t)-\hH_{{\rm kin},-}(t)],
 } to eliminate terms changing the number of doublons and holons in $\hH'^{(1)}(t)$. 
 We then obtain
 \eqq{
 \hH'^{(1)}(t) & =    \hH_{\rm kin, H}(t) +  \hH_{\rm kin, D}(t).
 }
 The second-order term can be expressed as 
  \eqq{
  \hH'^{(2)}(t) 
  =& \, i[\hS^{(1)}(t),\hH_{\rm kin, H}(t) +  \hH_{\rm kin, D}(t)] + i[\hS^{(2)}(t),\hH_U] \nonumber\\
  & +  \frac{1}{U} [\hH_{\rm kin,+}(t),\hH_{\rm kin,-}(t)].
 }
 Here, the first term changes the number of doublons and holons, but we can choose $\hS^{(2)}(t)$
 such that it is canceled by the second term.
 Therefore, the relevant contribution comes only from the third term, which consists of the two-site terms ($\hH_{\rm 2site}$) and the three-site terms ($\hH_{\rm 3site}$)~\cite{Murakami2022}.
In particular, $\hH_{\rm 2site}$ can be expressed as $\hH_{\rm 2site} =  \hH_{\rm Ushift} +  \hH_{\rm spin} +  \hH_{\eta}$,
where 
 \eqq{
 \hH_{\rm Ushift}  = &
  J_{{\rm ex},x} \sum_{{\bm i}} (\hn_{{\bm i}\uparrow}-\frac{1}{2})(\hn_{{\bm i}\downarrow}-\frac{1}{2}) \nonumber \\
& + \frac{J_{{\rm ex},y}}{2} \sum_{{\bm i}} (\hn_{{\bm i}\uparrow}-\frac{1}{2})(\hn_{{\bm i}\downarrow}-\frac{1}{2}), \\
 \hH_{\rm spin} =& J_{{\rm ex},x} \sum_{{\bm i}} \hat{\bf  s}_{{\bm i}}\cdot \hat{\bf  s}_{{\bm i}+{\bm e}_x} + J_{{\rm ex},y} \sum_{i_x} \hat{\bf  s}_{i_x0}\cdot \hat{\bf  s}_{i_x1} \label{eq:H_spin}, \\
  \hH_{\eta}  = & -\frac{J_{{\rm ex},x}}{2}\sum_{{\bm i}} \{e^{-2iA_x(t)}\heta_{{\bm i}}^+ \heta_{{\bm i}+{\bm e}_x}^- + e^{2iA_x(t)}\heta_{{\bm i}}^- \heta_{{\bm i}+{\bm e}_x}^+ \nonumber 
  \\ &+2\heta^z_{{\bm i}}\heta^z_{{\bm i}+{\bm e}_x} \} 
   -J_{{\rm ex},y}\sum_{i_x} \hat{\boldsymbol{\eta}}_{i_x0} \cdot \hat{\boldsymbol{\eta}}_{i_x1},
 }
  with $J_{{\rm ex},x}\equiv \frac{4t_x^2}{U}$ and $J_{{\rm ex},y}\equiv \frac{4t_y^2}{U}$. 
  We introduced the $\eta$-operators as $\heta^+_{{\bm i}} =(-1)^{i_x+i_y}\hc^\dagger_{{\bm i}\downarrow}  \hc^\dagger_{{\bm i}\uparrow}$, $\heta^-_{{\bm i}} = (\heta^+_{{\bm i}})^\dagger$ and $\heta^z_{{\bm i}} =\frac{1}{2}(\hn_{{\bm i}}-1)$~\cite{Yang1989PRL}.
  $\hH_{\rm Ushift}$ corresponds to the modification of the local Coulomb interaction, $\hH_{\rm spin}$ to the spin exchange coupling,
  and $\hH_{\eta}$ to the exchange coupling for neighboring doublons and holons. 
$\hH_{\rm 3site}$ represents the correlated hopping of a doublon and/or a holon. 
  If we focus on the hole-doped case and ignore $\hH_{\rm 3site}$, we obtain the celebrated $t-J$ model~\cite{Dagotto1994RMP}.
  
 Next, we consider the contribution from  $i(\partial_t e^{i\hS(t)}) e^{-i\hS(t)}$.
  At the order of $t_{xy}/U$, we have 
  \eqq{
  -\partial_t \hS^{(1)}(t) = -E_x(t)\hP_x(t),
  }
  where 
  \eqq{
  \hP_x(t) = \frac{i}{U} \frac{\partial}{\partial A_x(t)} [\hH_{{\rm kin},+}(t)-\hH_{{\rm kin},-}(t)]
  }
  can be regarded as the interband (doublon-holon) polarization.
  The explicit expression for $\hP_x(t)$ is 
  \eqq{
    \hP_x(t)  =& - \frac{t_x}{U} \sum_{{\bm i},\sigma} e^{-iA_x(t)} \hn_{{\bm i}\bar{\sigma}} \hc^\dagger_{{\bm i}\sigma}\hc_{{\bm i}+{\bm e}_x\sigma}  \hat{\bar{n}}_{{\bm i}+{\bm e}_x\bar{\sigma}}  \nonumber \\
    &+\frac{t_x}{U} \sum_{{\bm i},\sigma} e^{iA_x(t)} \hn_{{\bm i}+{\bm e}_x\bar{\sigma}}  \hc^\dagger_{{\bm i}+{\bm e}_x\sigma}\hc_{{\bm i}\sigma}  \hat{\bar{n}}_{{\bm i}\bar{\sigma}} \nonumber \\
    &+\frac{t_x}{U} \sum_{{\bm i},\sigma} e^{-iA_x(t)} \hat{\bar{n}}_{{\bm i}\bar{\sigma}} \hc^\dagger_{{\bm i}\sigma}\hc_{{\bm i}+{\bm e}_x\sigma}  \hn_{{\bm i}+{\bm e}_x\bar{\sigma}}  \nonumber \\
   & -\frac{t_x}{U} \sum_{{\bm i},\sigma} e^{iA_x(t)} \hat{\bar{n}}_{{\bm i}+{\bm e}_x\bar{\sigma}} \hc^\dagger_{{\bm i}+{\bm e}_x\sigma}\hc_{{\bm i}\sigma}  \hn_{{\bm i}\bar{\sigma}}.
  }

\begin{figure}[t]
  \centering
    \hspace{-0.cm}
    \vspace{0.0cm}
\includegraphics[width=85mm]{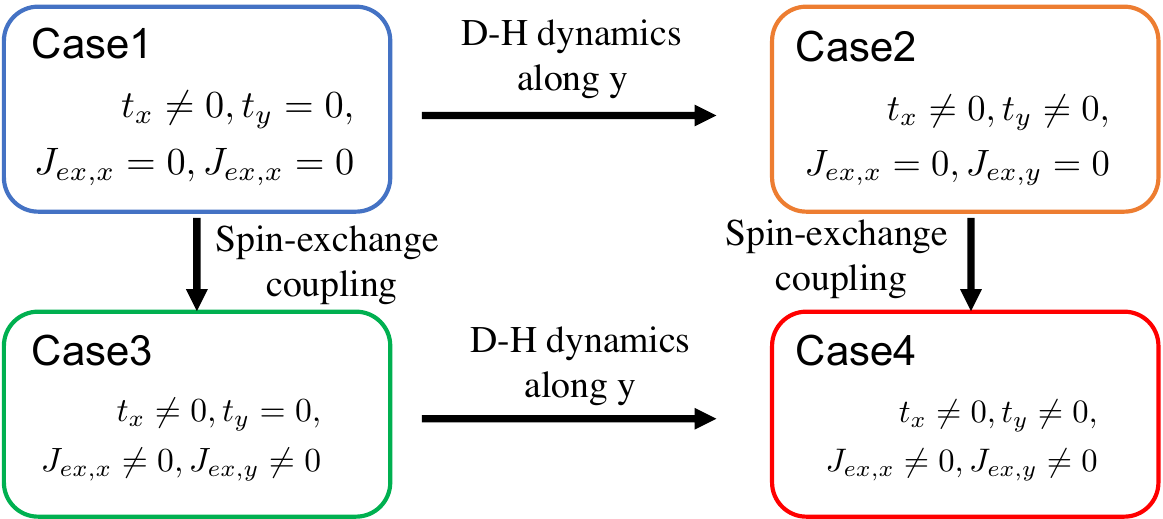} 
  \caption{ Illustration of the connection between the four cases considered in the effective model analysis.
  The arrows and attached comments indicate the physical processes added by switching from one case to the other.
}
  \label{fig:SW_model}
\end{figure}

In this work, in order to reveal the effects of the hopping of doublons and holons along the $y$ direction and the spin-charge coupling caused by the spin exchange coupling,
we ignore $ \hH_{\rm Ushift}$, $ \hH_{\eta}$, and $\hH_{\rm 3site}$ and focus on the effective model
\eqq{
\hH_{\rm eff}(t) = \hH_{\rm DH}(t) + \hH_{\rm spin} + \hH_U- E_x(t) \hat{P}_x(t).
}
Here $\hH_{\rm DH} = \hH_{\rm kin,D} + \hH_{\rm kin, H}$.
As shown below, this model successfully captures the qualitative changes in the HHG features between the models for zero and nonzero $t_y$. 
We simulate the light-induced dynamics using $\hH_{\rm eff}$ with some terms turned off or on, and evaluate the current~\eqref{eq:current} to obtain the HHG spectrum.
We prepare the ground state at half-filling using $\hH_{\rm spin}$ and consider the time evolution for the four cases shown in Fig.~\ref{fig:SW_model}.
Note that all these calculations
share the same ground state. 
The comparison between the different cases allows us to pinpoint the effects of different physical processes, as depicted in Fig.~\ref{fig:SW_model}.

\section{Perturbation theory from the strong-rung limit}
In this section, we introduce the perturbation theory from the strong-rung limit 
to extract information on the relevant elementary excitations in this regime~\cite{Endres1996PRB,Feiguin2019PRB}.
To this end, we first separate the model \eqref{eq:Hubbard_2leg} into
\eqq{
\hH_0 &= - t_y \sum_{i_x,\sigma} (\hc^\dagger_{i_x0\sigma} \hc_{i_x1\sigma} + h.c.) + \hH_U-\mu \sum_{{\bm i}} \hn_{{\bm i}} \nonumber \\
& \equiv \sum_{i_x} \hH_{{\rm rung},i_x} ,\\
\hH_I &= -t_x \sum_{{\bm i},\sigma} (\hc^\dagger_{{\bm i}+{\bm e}_x\sigma} \hc_{{\bm i}\sigma} + h.c.) \nonumber \\
& \equiv\sum_{i_x} \hH_{I}^{[i_x,i_x+1]}. \label{eq:H_I}
}
We assume that the energy scale of $\hH_0$ dominates $\hH_I$ and treat $\hH_I$ perturbatively. 
In the following, we set $\mu=U/2$, assuming the half-filling condition.

\subsection{Eigenstates and eigenvalues of $\hH_{{\rm rung},i}$}
At the 0th order, the eigenstates correspond to those of $\hH_0$.
Each eigenstate can be expressed as a direct product of an eigenstate at each rung $\hH_{{\rm rung},i_x}$.

\begin{figure*}[t]
     \centering
\includegraphics[width=160mm]{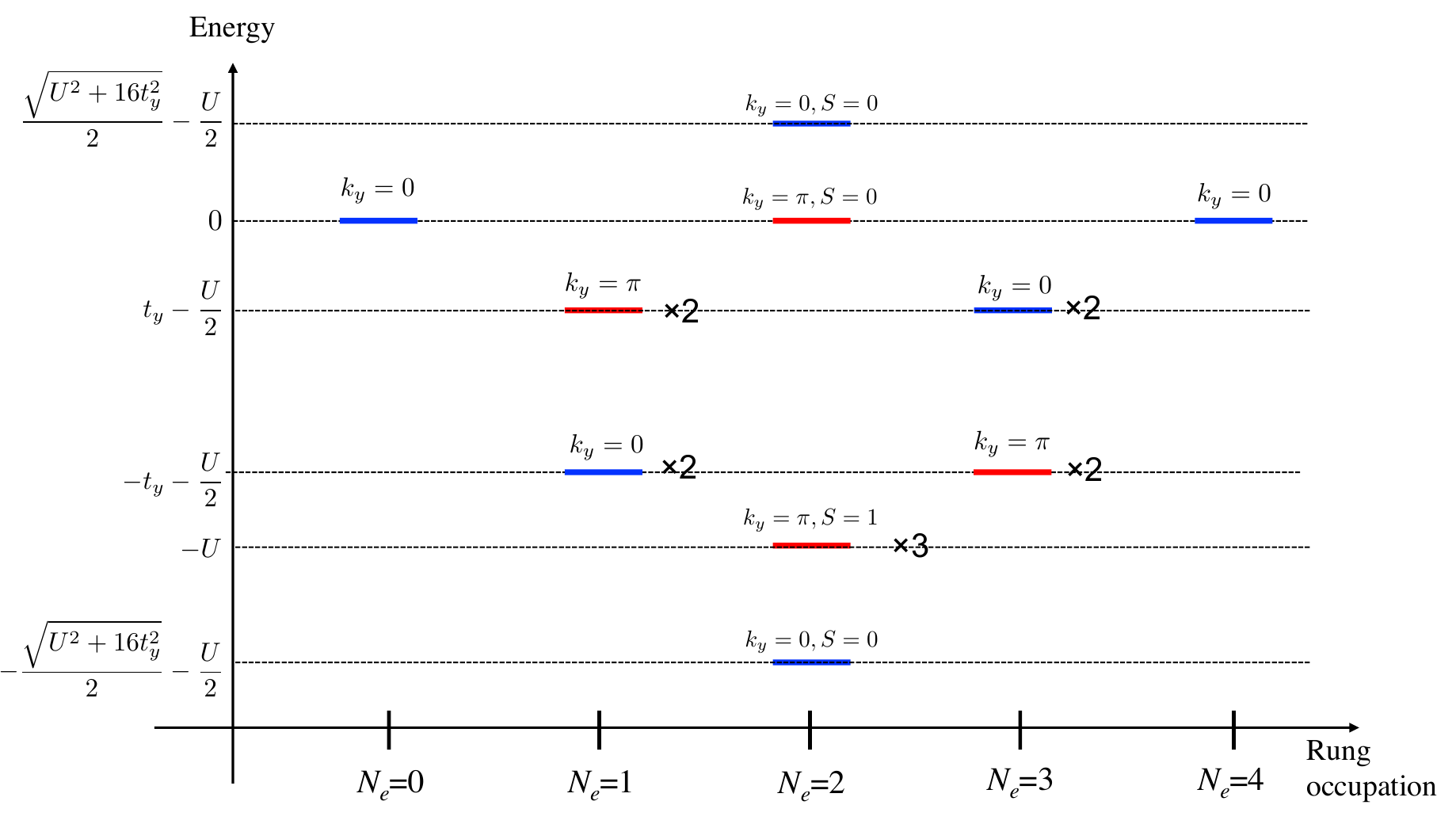}
\caption{Schematic picture of the distribution of the eigenstates on a rung, in the plane of the electron number ($N_e$) and the energy.
The symbols $\times2$ and $\times3$ indicate the degree of degeneracy.
}
\label{fig:E_rung}
\end{figure*}

We summarize the eigenstates on a given rung in Tab.~\ref{tab:rung} and Fig.~\ref{fig:E_rung}.
For simplicity, we omit the site index, $i_x$, but explicitly keep the chain index $i_y(=0,1)$. 
The eigenstates are categorized by the total electron number $N_{e}$ on the rung and the momentum along the $y$ direction $k_y$ (or equivalently the parity for
the mirror operation along the $y$ direction). Since we have two sites along $y$, $k_y$ is either $0$ (parity even) or $\pi$ (parity odd). 
For $|N_e=2,k_y=0,S=0,-\rangle$ and $|N_e=2,k_y=0,S=0,+\rangle$ in the table, we introduced the basis set 
\eqq{
|X \rangle &=   \frac{1}{\sqrt{2}} (\hc^\dagger_{0\uparrow}\hc^\dagger_{0\downarrow} + \hc^\dagger_{1\uparrow}\hc^\dagger_{1\downarrow})|{\rm vac}\rangle, \nonumber \\
|Y \rangle &=   \frac{1}{\sqrt{2}} (\hc^\dagger_{0\uparrow}\hc^\dagger_{1\downarrow}- \hc^\dagger_{0\downarrow}\hc^\dagger_{1\uparrow})|{\rm vac}\rangle.
}
The corresponding eigenvalues are 
\eqq{
\mathcal{E}_{\mp} = \mp \frac{\sqrt{U^2+16t_y^2}}{2} -\frac{U}{2},
}
and the coefficients of the eigenstates in the basis of $\{|X\rangle,|Y\rangle\}$ are 
\eqq{
\cos \frac{\theta}{2} &= \frac{1}{\sqrt{2}} \sqrt{1+\frac{U}{\sqrt{U^2 + 16t_y^2}}}, \nonumber \\
\sin \frac{\theta}{2}  &= -\frac{1}{\sqrt{2}} \sqrt{1-\frac{U}{\sqrt{U^2 + 16t_y^2}}},
}
see Tab.~\ref{tab:rung}.

\begin{table*}[tbh]
\centering
\begin{tabular}{|c|c|c|c|c|c|}
\hline 
$N_e$ & $k_y$ & Name & Expression & Energy & Degeneracy \\
\hline \hline
0 & 0 & $|N_e=0\rangle$ & $|{\rm vac}\rangle$ & 0 & 1 \\
\hline
1 & 0 & $|N_e=1,k_y=0,\sigma\rangle$ & $ \frac{1}{\sqrt{2}} (\hc^\dagger_{0\sigma} + \hc^\dagger_{1\sigma})|{\rm vac}\rangle$ & $-t_y -\frac{U}{2}$ & 2 \\
\hline
1 & $\pi$ & $|N_e=1,k_y=\pi,\sigma\rangle$ & $ \frac{1}{\sqrt{2}} (\hc^\dagger_{0\sigma} - \hc^\dagger_{1\sigma})|{\rm vac}\rangle $ &  $t_y -\frac{U}{2}$ & 2 \\
\hline
2 & $\pi$ & $|N_e=2,k_y=\pi,S=0\rangle$ & $ \frac{1}{\sqrt{2}} (\hc^\dagger_{0\uparrow}\hc^\dagger_{0\downarrow}- \hc^\dagger_{1\uparrow}\hc^\dagger_{1\downarrow})|{\rm vac} \rangle $ & $0$ & 1 \\
\hline
2 & $\pi$ & \makecell{$|N_e=2,k_y=\pi,S=1, S_z=1\rangle$ \\ $|N_e=2,k_y=\pi,S=1, S_z=-1\rangle$ \\$|N_e=2,k_y=\pi,S=1, S_z=0\rangle$} & 
\makecell{ $\hc^\dagger_{0\uparrow}\hc^\dagger_{1\uparrow}|{\rm vac}\rangle$\\ $\hc^\dagger_{0\downarrow}\hc^\dagger_{1\downarrow}|{\rm vac}\rangle$ \\ $ \frac{1}{\sqrt{2}} (\hc^\dagger_{0\uparrow}\hc^\dagger_{1\downarrow}+ \hc^\dagger_{0\downarrow}\hc^\dagger_{1\uparrow})|{\rm vac} \rangle $}& $-U$ & 3 \\
\hline
2 & 0 & $|N_e=2,k_y=0,S=0,-\rangle$ & $ -\sin \frac{\theta}{2} |X\rangle +\cos \frac{\theta}{2} |Y\rangle $ & $\mathcal{E}_-$ & 1 \\
\hline
2 & 0 & $|N_e=2,k_y=0,S=0,+\rangle$ & $ \cos \frac{\theta}{2} |X\rangle +\sin \frac{\theta}{2} |Y\rangle  $ & $\mathcal{E}_+$ & 1 \\
\hline
3 & 0 & $|N_e=3,k_y=0,\sigma\rangle$ & $ \frac{1}{\sqrt{2}} (\hc_{0\sigma} + \hc_{1\sigma})|N=4\rangle$ & $t_y -\frac{U}{2}$ & 2 \\
\hline
3 & $\pi$ & $|N_e=3,k_y=\pi,\sigma\rangle$ & $ \frac{1}{\sqrt{2}} (\hc_{0\sigma} - \hc_{1\sigma})|N=4\rangle $ &  $-t_y -\frac{U}{2}$ & 2 \\
\hline
4 & 0 & $|N_e=4\rangle$ & $\hc^\dagger_{0\uparrow}\hc^\dagger_{0\downarrow}\hc^\dagger_{1\uparrow}\hc^\dagger_{1\downarrow} |{\rm vac}\rangle $ & 0 & 1  \\
\hline
\end{tabular}
\caption{List of eigenstates on a rung, categorized according to the electron number ($N_e$) and the momentum along the $y$ direction ($k_y$). For $|N_e=2,k_y=0,S=0,-\rangle$ and $|N_e=2,k_y=0,S=0,+\rangle$, we use $|X\rangle$, $|Y\rangle$, $\theta$, $\mathcal{E}_{-}$ and $\mathcal{E}_{+}$ defined in the text. }
\label{tab:rung}
\end{table*}

\subsection{Perturbation analysis} \label{sec:rung_perturb}
An eigenstate of $\hH_0$ can be expressed as a direct product of eigenstates for each rung.
Using the singlet state $|S\rangle\equiv |N_e=2,k_y=0,S=0,- \rangle$,
the ground state for $\hH_0$ can be expressed as 
\eqq{
|\Psi_0\rangle = |S\rangle_0 |S\rangle_1\cdots |S\rangle_{N_x-1}. \label{eq:GS_rung}
}
In the following, we apply degenerate perturbation theory 
to describe the dispersion of polarons and magnons.

\subsubsection{Polarons}
In this section, we consider states where we add or remove one electron from the ground state
to describe polarons.
First, we focus on the degenerate Hilbert space consisting of states with one of the $|S\rangle$'s in the ground state $|\Psi_0\rangle$ (Eq.~\eqref{eq:GS_rung}) replaced by $|N_e=1,k_y,\sigma\rangle$.
We denote the state where $|N_e=1,k_y,\sigma\rangle$ is introduced at site $l$ by $|\Psi_{\rm hP}(l,k_y,\sigma)\rangle$.
These states represent a bound state of the unpaired fermion and a hole~\cite{Feiguin2019PRB}. Since they host a positive charge $-q$, relative to the ground state, as well as a spin degree of freedom, they provide a basis for the hole-like polaron, which we call 
hPolaron. 

Since $\hH_I$ (Eq.~\eqref{eq:H_I}) conserves the momentum $k_y$ (the parity along $y$, $P_y$) and total $S_z$, 
we can separately apply degenerate perturbation theory for the four cases $\{k_y=0,\pi,\sigma=\uparrow,\downarrow\}$.
We have 
\eqq{
& \langle \Psi_{\rm hP}(i,k_y,\sigma)| \hH_I |\Psi_{\rm hP}(j,k_y,\sigma)\rangle =\nonumber \\
& \delta_{|i-j|,1}\times
 \left[t_x \frac{\{4t_y - e^{ik_y}(U -\sqrt{16t_y^2 + U^2})\}^2}{64t_y^2 + 4U(U-\sqrt{16t_y^2+U^2})} \right].
}
Because of the translational invariance of $\langle \Psi_{\rm hP}(i,k_y,\sigma)| \hH_I |\Psi_{\rm hP}(j,k_y,\sigma)\rangle$ with respect to $i$ and $j$, 
the eigenstates have the form 
\eqq{
|\Psi_{\rm hP}(k_x,k_y,\sigma)\rangle = \sum_l e^{ik_xl}|\Psi_{\rm hP}(l,k_y,\sigma)\rangle  \label{eq:wf_hP}
}
and the corresponding energy is 
\eqq{
\mathcal{E}^{(1)}_{\rm hP}(k_x,k_y) = t_x A(k_y) \cos(k_x).
}
Here
\eqq{
A(k_y)  =  \frac{\{4t_y - e^{ik_y}(U -\sqrt{16t_y^2 + U^2})\}^2}{32t_y^2 + 2U(U-\sqrt{16t_y^2+U^2})}.
}
Note that $\mathcal{E}^{(1)}_{\rm hP}$ describes the dispersion of the hPolaron which incorporates the first-order correction of $t_x$.

Next we focus on the degenerate Hilbert space consisting of states with one of the $|S\rangle$'s in $|\Psi_0\rangle$ being replaced by $|N_e=3,k_y,\sigma\rangle$.
The state where $|N_e=3,k_y,\sigma\rangle$ is introduced at site $l$ is denoted by $|\Psi_{\rm eP}(l,k_y,\sigma)\rangle$.
Since these states host a negative charge $q$ compared to the ground state as well as a spin degree of freedom, they provide a basis for the electron-like polaron, which we denote by
ePolaron. 
As in the above case, we have 
\eqq{
& \langle \Psi_{\rm eP}(i,k_y,\sigma)| \hH_I |\Psi_{\rm eP}(j,k_y,\sigma)\rangle =  \nonumber \\
&\delta_{|i-j|,1}\times
 \Biggl[-t_x \frac{\{4t_y + e^{ik_y}(U -\sqrt{16t_y^2 + U^2})\}^2}{64t_y^2 + 4U(U-\sqrt{16t_y^2+U^2})} \Biggl].
}
The eigenstates can be expressed as 
\eqq{
|\Psi_{\rm eP}(k_x,k_y,\sigma)\rangle = \sum_l e^{ik_xl}|\Psi_{\rm eP}(l,k_y,\sigma)\rangle, 
}
and the corresponding energies are
\eqq{
\mathcal{E}^{(1)}_{\rm eP}(k_x,k_y) = -t_x A'(k_y) \cos(k_x) 
}
with 
\eqq{
A'(k_y)  =  \frac{\{4t_y + e^{ik_y}(U -\sqrt{16t_y^2 + U^2})\}^2}{32t_y^2 + 2U(U-\sqrt{16t_y^2+U^2})} .
}

To summarize, the energy of the polarons, measured from the ground state, is 
\eqq{
\mathcal{E}_{\rm hP}(k_x,k_y) &=   t_x A(k_y) \cos(k_x)  -e^{ik_y}t_y +\frac{\sqrt{16t_y^2+U^2}}{2}, \label{eq:hole_polaron} \\
\mathcal{E}_{\rm eP}(k_x,k_y) &= -t_x A'(k_y) \cos(k_x) +e^{ik_y}t_y +\frac{\sqrt{16t_y^2+U^2}}{2}. \label{eq:electron_polaron}
}
We note that these are the dispersions expected to be observed in the single-particle spectrum 
if the polarons can be created by adding an electron to or removing an electron from the ground state. 

\subsubsection{Magnons}
Now we focus on the degenerate Hilbert space consisting of states with one of the $|S\rangle$'s in $|\Psi_0\rangle$ replaced by a triplet state $|M_{S_z}\rangle$. 
Here  $|M_{1}\rangle=|N_e=2,k_y=\pi, S=1, S_z=1\rangle, |M_{-1}\rangle=|N=2,k_y=\pi, S=1,S_z=-1\rangle$ and $|M_{0}\rangle=|N=2,k_y=\pi, S=1, S_z=0\rangle$.
We denote a state where  $|M_{S_z}\rangle$ is introduced at site $l$ by $|\Psi_{\rm M}(l,k_y=\pi,S_z)\rangle$.
Note that these states host a spin degree of freedom, but they have no charge.
We now apply the second-order degenerate perturbation theory, where the matrix elements of the effective Hamiltonian are expressed as 
\eqq{
\langle \alpha|\hH^{(2)}_{\rm eff}|\beta\rangle=\sum_n \frac{\langle\alpha|\hH_I|n\rangle \langle n| \hH_I|\beta\rangle}{\mathcal{E}_0-\mathcal{E}_n}.
}
Here $\hH^{(2)}_{\rm eff}$ represents the effective model in the degenerate subspace at the order of $\mathcal{O}(t_x^2)$.
 $|\alpha\rangle,|\beta\rangle$ are orthogonal states in the degenerate subspace with energy $\mathcal{E}_0$ and $|n\rangle$ is an eigenstate outside this subspace, whose energy is $\mathcal{E}_n$. Due to the conservation of $\hS_z$, we can treat the cases of $S_z = -1,0,1$, separately.

For $|\beta\rangle=|\Psi_{\rm M}(l,k_y=\pi,S_z)\rangle$, $\langle \alpha|\hH^{(2)}_{\rm eff}|\beta\rangle$ is nonzero only if $|\alpha\rangle=|\Psi_{\rm M}(l-1,k_y=\pi,S_z)\rangle,|\Psi_{\rm M}(l,k_y=\pi,S_z)\rangle,|\Psi_{\rm M}(l+1,k_y=\pi,S_z)\rangle$. In addition, the elements for $(l+1,l)$ and $(l-1,l)$ are the same.
The off-diagonal element can be expressed as 
\eqq{
& \langle \Psi_{\rm M}(l-1,k_y=\pi,S_z)|\hH^{(2)}_{\rm eff}|\Psi_{\rm M}(l,k_y=\pi,S_z)\rangle \nonumber \\
&=2t_x^2 \Bigl(\frac{2}{U}-\frac{1}{\sqrt{16t_y^2+U^2}} \Bigl)\equiv B(t_x,t_y,U).
}
The diagonal element is 
\eqq{
\langle \beta |\hH^{(2)}_{\rm eff}|\beta\rangle 
& = \sum_n \frac{\langle\alpha|\hH^{[l,l+1]}_I|n\rangle \langle n| \hH^{[l,l+1]}_I|\alpha\rangle}{\mathcal{E}_0-\mathcal{E}_n} \nonumber 
\\&+ \sum_n \frac{\langle\alpha|\hH^{[l-1,l]}_I|n\rangle \langle n| \hH^{[l-1,l]}_I|\alpha\rangle}{\mathcal{E}_0-\mathcal{E}_n}  \label{eq:H_magnon}
\\&+\sum_{i\neq l,l-1} \sum_n \frac{\langle\alpha|\hH^{[i,i+1]}_I|n\rangle \langle n| \hH^{[i,i+1]}_I|\alpha\rangle}{\mathcal{E}_0-\mathcal{E}_n},\nonumber  
}
where 
\eqq{
 &\sum_n \frac{\langle\alpha|\hH^{[l,l+1]}_I|n\rangle \langle n| \hH^{[l,l+1]}_I|\alpha\rangle}{\mathcal{E}_0-\mathcal{E}_n} \nonumber \\
& =\sum_n \frac{\langle\alpha|\hH^{[l-1,l]}_I|n\rangle \langle n| \hH^{[l-1,l]}_I|\alpha\rangle}{\mathcal{E}_0-\mathcal{E}_n} =-B(t_x,t_y,U),
}
 and for $i\neq l,l-1$
\eqq{
& \sum_n \frac{\langle\alpha|\hH^{[i,i+1]}_I|n\rangle \langle n| \hH^{[i,i+1]}_I|\alpha\rangle}{\mathcal{E}_0-\mathcal{E}_n} \nonumber \\&= -2t_x^2 \frac{U^2}{(16t_y^2+U^2)^{3/2}} \equiv C(t_x,t_y,U). 
\label{eq:C_term}
}

We note that the contribution~\eqref{eq:H_magnon} appears when considering the second-order energy correction to the ground state.
Due to the translational invariance of the system, we can construct the wave function of the magnon as a plane wave consisting of the local state
as in Eq.~\eqref{eq:wf_hP} for the polarons.
Considering these points, we obtain the energy of the magnon measured from the ground state as 
\begin{align}
&\mathcal{E}_{\rm M}(k_x,k_y=\pi)= 2B(t_x,t_y,U) \cos(k_x) -2B(t_x,t_y,U) \nonumber \\
&\quad-2C(t_x,t_y,U) + \frac{1}{2}\Bigl(-U+\sqrt{U^2+16t_y^2}\Bigl). \label{eq:magnon}
\end{align}

\subsubsection{Magonon-polaron continuum and spin-bag state}
The single-particle spectrum $A(\bk,\omega)$ represents eigenstates accessed by adding an electron to or removing an electron from a system~\cite{Fetter2003,ARPES_review}.
In many-body systems, one electron may consist of a few elementary excitations. In such a case, a continuum emerges in the single-particle spectrum.
In the Mott insulator on the two-leg ladder, we see a continuum consisting of a magnon (Eq.~\eqref{eq:magnon}) and a polaron (Eq.~\eqref{eq:hole_polaron} or Eq.~\eqref{eq:electron_polaron}).

For simplicity, in this section, we write the magnon dispersion (Eq.~\eqref{eq:magnon}) as $\mathcal{E}_{\rm M}(k_x) = X \cos(k_x) + Y$, and the polaron dispersion (Eq.~\eqref{eq:hole_polaron} or Eq.~\eqref{eq:electron_polaron}) as $\mathcal{E}_{\rm P}(k_x) = Z \cos (k_x) + W$. If the interaction between the magnon and the polaron is sufficiently weak, a state hosting one magnon and one polaron is characterized by the total momentum along $x$ ($k_{\rm tot}$) and the momentum of the magnon along $x$ ($k$). The corresponding energy can be expressed as 
\eqq{
\mathcal{E}_{\rm MP} (k_{\rm tot},k) = \mathcal{E}_{\rm M}(k) + \mathcal{E}_{\rm P}(k_{\rm tot}-k).
}
It is easy to see that, by changing $k\in [-\pi,\pi)$ and fixing $k_{\rm tot}$, that $\mathcal{E}_{\rm MP} (k_{\rm tot},k)$ covers the range 
\eqq{
&Y+W- \sqrt{X^2+Z^2 + 2XZ\cos(k_{\rm tot})} \leq \mathcal{E}_{\rm MP} (k_{\rm tot},k) \nonumber \\
 &\;\;\;\;\; \leq Y+W+ \sqrt{X^2+Z^2 + 2XZ\cos(k_{\rm tot})} . \label{eq:contunum}
}

As we will see below in Fig.~\ref{fig:Akw_ty2}, the single-particle spectrum $A(k_x,k_y=0,\omega)$ for the half-filled two-leg Hubbard model 
shows a continuum consisting of a magnon and an ePolaron with $k_y=\pi$.
The intensity at the bottom of the continuum is particularly strong. 
It has been attributed to a loosely bound state of a magnon-polaron pair, a new composite state called spin-bag~\cite{Feiguin2019PRB}.
In the following, we thus denote the bottom edge of Eq.~\eqref{eq:contunum} for a magnon and an ePolaron with $k_y=\pi$  by $\mathcal{E}_{\rm spin-bag}(k_x,k_y=0)$.

\section{Semiclassical three-step model}
In this section, we briefly summarize the idea of the semiclassical three-step model.~\cite{Lewenstein1994,Vampa2015PRB,Murakami2021PRB}.
In this model, we focus on the relative dynamics of an elementary excitation with a positive charge $-q$ and that with a negative charge $q$.
For example, in conventional semiconductors, the former is a hole and the latter is an electron~\cite{Vampa2015PRB}.
The three-step picture encompasses of the following steps: (i) the pair of elementary excitations is created at time $t_{\rm cr}$ by tunneling at the gap minimum in momentum space and spatially at the same point,
(ii) they move around following a semiclassical kinetic equation, and (iii) they return to the original (relative) position and recombine to emit the energy.
The relative dynamics is described for the one-dimensional system by the equations 
\eqq{
\frac{d x_{\rm rel}(t)}{dt} &= \frac{d \mathcal{E}_{\rm g}(k_x)}{dk_x} \Bigl|_{k_x=k_x(t)}, \\
k_x(t) &= k_0 -qA_x(t) + qA_x(t_{\rm cr}).
} 
Here $x_{\rm rel}(t)$ denotes the relative distance of the two elementary excitations, where $x_{\rm rel}(t_{\rm cr}) = 0$.
$\mathcal{E}_{\rm g}(k_x)$ is the energy of a pair with zero total momentum, where $k_x$ denotes the momentum of the elementary excitation with charge $q$.
Note that only the pair with zero total momentum can be excited with the homogeneous electric field.
For semiconductors, we have  $\mathcal{E}_{\rm g}(k_x)=\mathcal{E}_{\rm c}(k_x)-\mathcal{E}_{\rm v}(k_x)$, where $\mathcal{E}_{\rm c}(k_x)$ ($\mathcal{E}_{\rm v}(k_x)$) denotes the dispersion of the conduction (valence) band. 
$k_0$ is the location of the gap minimum.
The recombination occurs at $t_{\rm rec}(>t_{\rm cr})$ when $x_{\rm rel}(t_{\rm rec}) = 0$, and the frequency of the emitted light is equal to the energy of the pair at that time, i.e. $\omega_{\rm emit}(t_{\rm rec})=\mathcal{E}_{\rm g}(k_x(t_{\rm rec}))$.

As numerically shown in Ref.~\cite{Murakami2021PRB} for the 1D single-chain Hubbard model, this description can also be applied to strongly correlated systems, provided that we use the proper information of elementary excitations~\cite{Imai2022PRR}.
In the following, we present the explicit expression of $\mathcal{E}_{\rm g}(k_x)$ for different scenarios of HHG in the Mott insulator on the two-leg ladder.

\subsection{Doublon-holon pair}
In the 1D single-chain Hubbard model ($t_y=0$), the dispersion relation of the elementary excitations is exactly known~\cite{Essler2005}.
In particular, a holon (doublon) is parametrized by a quantity called rapidity $\xi$, where the corresponding momentum $k_H(\xi)$ ($k_D(\xi)$) and the energy $\mathcal{E}_H(\xi)$ ($\mathcal{E}_{\rm D}(\xi)$)
are given by
\begin{subequations}\label{eq:Bethe}
\eqq{
& k_H(\xi) = k_D(\xi) + \pi = \nonumber \\
&\;\;\; \frac{\pi}{2}-\xi-2\int_0^\infty \frac{d\omega}{\omega} \frac{\mathcal{J}_0(\omega) \sin(\omega\sin \xi)}{1+\exp(\frac{U\omega}{2})}, \\
&\mathcal{E}_{\rm H}(\xi) = \mathcal{E}_{\rm D}(\xi) =  \\
&\;\;\;\frac{U}{2}+2\cos \xi+2\int_0^\infty \frac{d\omega}{\omega} \frac{\mathcal{J}_1(\omega) \cos(\omega\sin \xi)e^{-\frac{U\omega}{4}}}{\cosh(\frac{U\omega}{4})},\nonumber
}
\end{subequations}
with $\xi \in \mathbb{R}$, respectively. Here $\mathcal{J}_n$ is the $n$th Bessel function.
We can obtain the energy as a function of momentum from these equations.
Then, we have 
\eqq{
\mathcal{E}_{\rm g}(k_x) = \mathcal{E}_{\rm H}(-k_x) + \mathcal{E}_{\rm D}(k_x).\label{eq:DH_pair}
}
In our analysis, we use Eq.~\eqref{eq:DH_pair} even for $t_y\neq 0$, but it successfully captures the relevant dynamics as can be seen in the subcycle spectra.

\subsection{Scenario 1 (hPolaron + ePolaron + magnon) in the strong-rung regime }
This scenario involves one hPolaron with $k_y=0$, one ePolaron with $k_y=\pi$ and one magnon with $k_y=\pi$.
The hPolaron carries the charge $-q$, the ePolaron carries the charge $q$, and the magnon carries no charge.
Therefore, the dynamics under the electric field involves only the former two elementary excitations, while the magnon provides an additional energy shift to $\mathcal{E}_{\rm g}(k_x)$.
Given that the minimum energy of a magnon is at $\bk=(\pi, \pi)$, we have 
\eqq{
\mathcal{E}_{\rm g}(k_x) = \mathcal{E}_{\rm hP}(-k_x-\pi,0) + \mathcal{E}_{\rm eP}(k_x,\pi) + \mathcal{E}_{\rm M}(\pi,\pi).
}
In practice, when we simulate the three-step model, we add a slight energy shift $\Delta \omega=0.4$, to be consistent with the numerically obtained single-particle spectrum, see Sec.~\ref{sec:Akw} below.
This shift can be attributed to higher order corrections missing in Eqs.~\eqref{eq:hole_polaron}, \eqref{eq:electron_polaron}, and \eqref{eq:magnon}.

\subsection{Scenario 2 (hPolaron + spin bag) in the strong-rung regime}
In this scenario, we assume that an ePolaron and a magnon in Scenario 1 move together as a weakly-bound composite particle, called a spin bag.
In this case, we have 
\eqq{
\mathcal{E}_{\rm g}(k_x) =\mathcal{E}_{\rm hP}(-k_x,0) + \mathcal{E}_{\rm spin-bag}(k_x,0).
}
In practice, when we simulate the three-step model, we add a slight energy shift $\Delta \omega=0.4$, to be consistent with the single-particle spectrum, see Sec.~\ref{sec:Akw} below.

\section{Supplementary results}
\begin{figure}[t]
     \centering
\includegraphics[width=70mm]{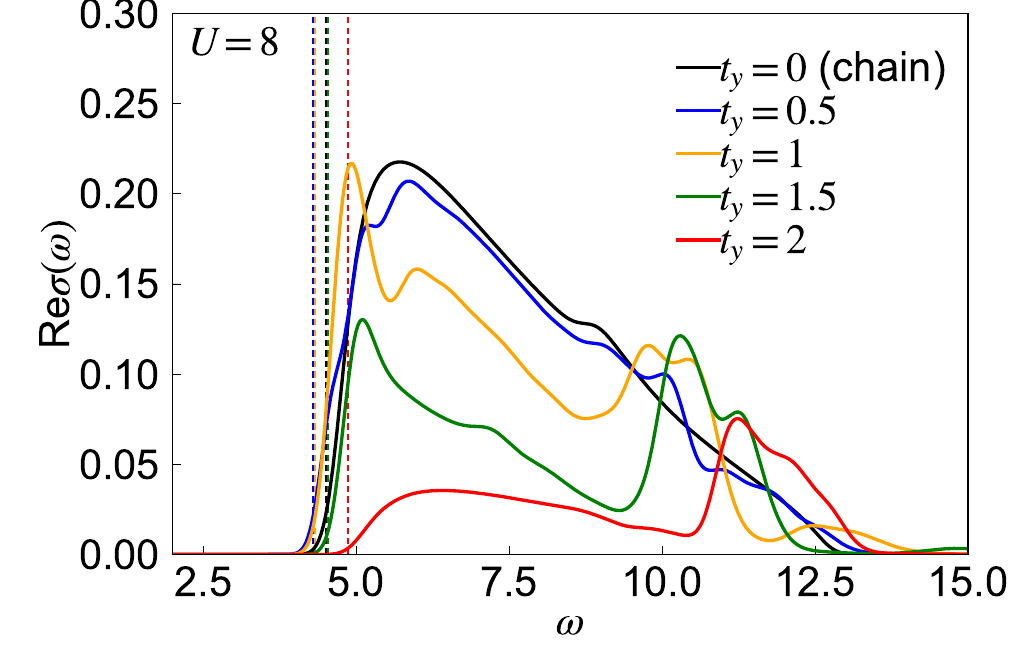}
\caption{Real part of the optical conductivity along $x$ (${\rm Re}\sigma(\omega)$) for the half-filled two-leg Hubbard model.
The system parameters are $t_x=1$ and $U=8$.
We use iTEBD and set  $\sigma_p=5$ for the Fourier transform.
The vertical dashed lines indicate the position of the gap, estimated as $\omega$ where ${\rm Re}\sigma(\omega)$ exceeds 10\% of the maximum intensity around the gap.  
For $t_y=0$, we prepare the equilibrium state with the cutoff dimension $D=1000$, while the time evolution is calculated with $D=3000$.
For nonzero $t_y$, we prepare the equilibrium state with $D=3000$, while the time evolution is calculated with $D=6000$.}
\label{fig:op_cond}
\end{figure}

\subsection{Optical conductivity}
To estimate the gap size, we evaluate the linear optical conductivity using iTEBD.
Namely, we apply a weak Gaussian field $E_{\rm probe}(t)$ along the $x$ direction, which is centered at a certain time $t_p$, and we measure the current $J_x(t)$.
We evaluate the optical conductivity as $J_x(\omega)/E_{\rm  probe}(\omega)$, where $J_x(\omega)$ and $E_{\rm  probe}(\omega)$ are the Fourier transforms of $J_x(t)$ and $E_{\rm proble}(t)$, respectively.
Since the numerical simulation is limited to finite times, we apply a Gaussian window $F_G(t,t_p) = \exp(-\frac{(t-t_p)^2}{2\sigma_p^2})$ to $J_x(t)$ in the Fourier transform, 
which leads to a broadening of the spectrum.

The results are shown in Fig.~\ref{fig:op_cond} for $\sigma_p=5$. For $t_y=1$, we reproduce the three-peak-like structure previously pointed out in Ref.~\cite{Shinjo2021PRB}.
For $t_y\lesssim 1.5$, the gap size is almost unchanged, while for $t_y=2$, the gap is clearly increased compared to the case of $t_y=0$.
We estimate the value of the gap from the point where ${\rm Re}\sigma(\omega)$ exceeds 10\% of the maximum intensity around the gap.

In a previous study on semiconductor HHG, the similarity between the shape of the HHG spectrum and that of the optical conductivity was pointed out~\cite{Yoshikawa2019}.
In the present system, when we compare Fig.~\ref{fig:op_cond} and Fig.~1 in the main text, it is hard to see the similarity. 
In particular, there is no signature in HHG corresponding to the peaks in the optical conductivity (see, for example, the peak around $\omega=10$ for $t_y=1$).

 \begin{figure}[t]
  \centering
    \hspace{-0.cm}
    \vspace{0.0cm}
\includegraphics[width=85mm]{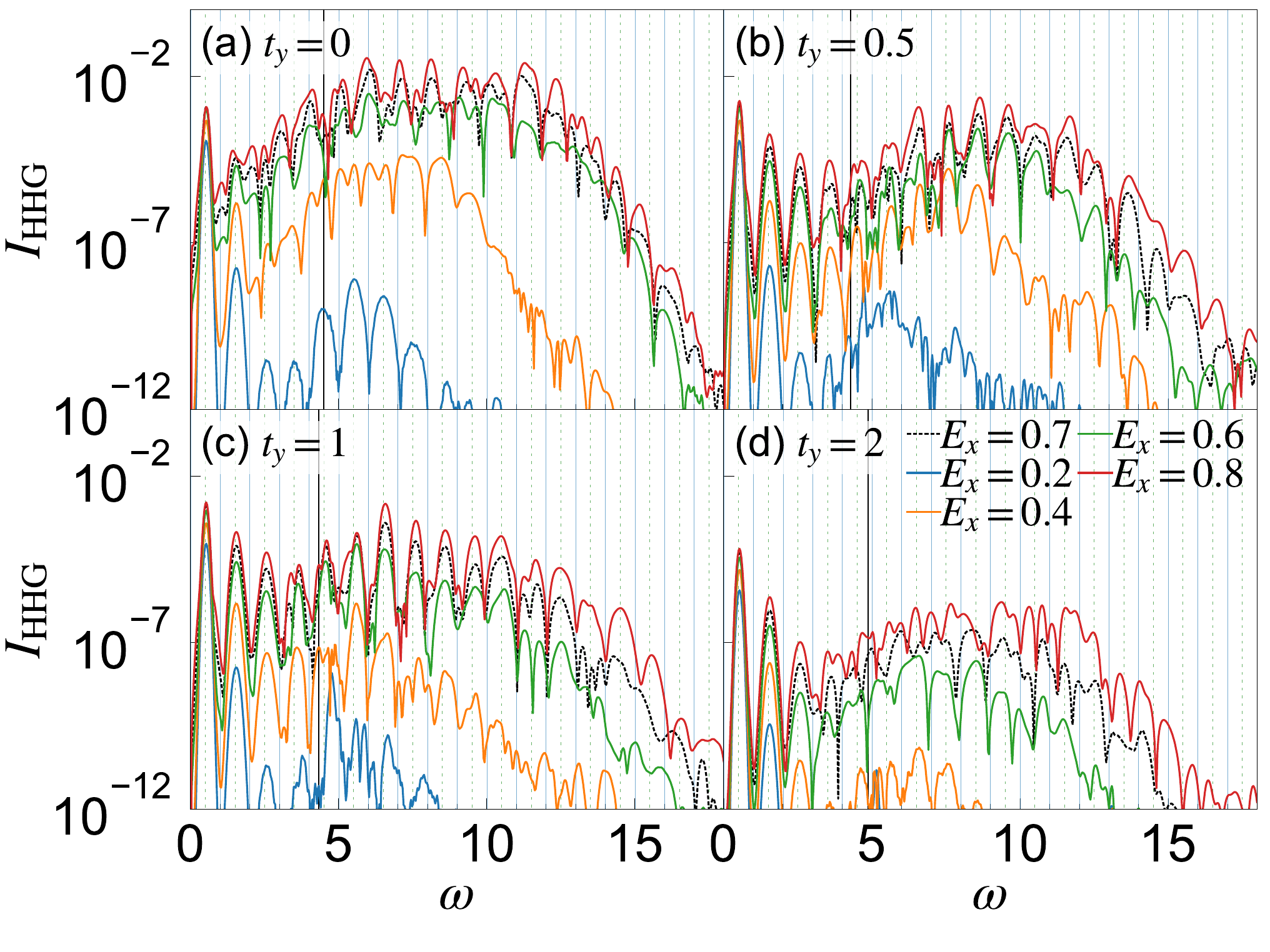} 
  \caption{HHG spectra for the two-leg Hubbard model at half filling for different values of $t_y$ and $E_x$.
  We use $t_x=1$ and $U=8$, and set the excitation parameters to $\Omega=0.5$, $t_0=60$ and $\sigma_0=15$.
  The black vertical lines indicate the optical gap.
}
  \label{fig:HHG_Ex_dep}
\end{figure}

\subsection{HHG spectrum and subcycle analysis}
In this section, we provide supplementary results for the HHG spectra and the subcycle analyses. 
In Fig.~\ref{fig:HHG_Ex_dep}, we show the dependence of the HHG spectra on $E_x$ for different values of $t_y$.
For $t_y=0$, i.e. the decoupled chains, one can essentially see a single plateau, whose cut-off frequency increases with increasing $E_x$.
As pointed out in the main text, in general, peak structures are not prominent and non-odd-harmonic signals ($\omega\neq (2n+1)\Omega$) appear despite the expectation from the inversion symmetry.
For $t_y=0.5$, peak structures become more pronounced and they develop at $\omega=(2n+1)\Omega$.
For $t_y=1$, peak structures remain pronounced in general. 
One can identify the development of a hump just above the gap ($4.5\lesssim\omega\lesssim6.5$), which leads to a plateau-like structure.
The position of the hump shifts to higher frequency with increasing $E_x$.
In the higher frequency part well above the gap, one can identify the development of another plateau-like structure, whose cut-off frequency also increases with increasing $E_x$.
For $t_y=2$, peak structures at $\omega=(2n+1)\Omega$ become less clear, as we pointed out in the main text.

 \begin{figure}[t]
  \centering
    \hspace{-0.cm}
    \vspace{0.0cm}
\includegraphics[width=70mm]{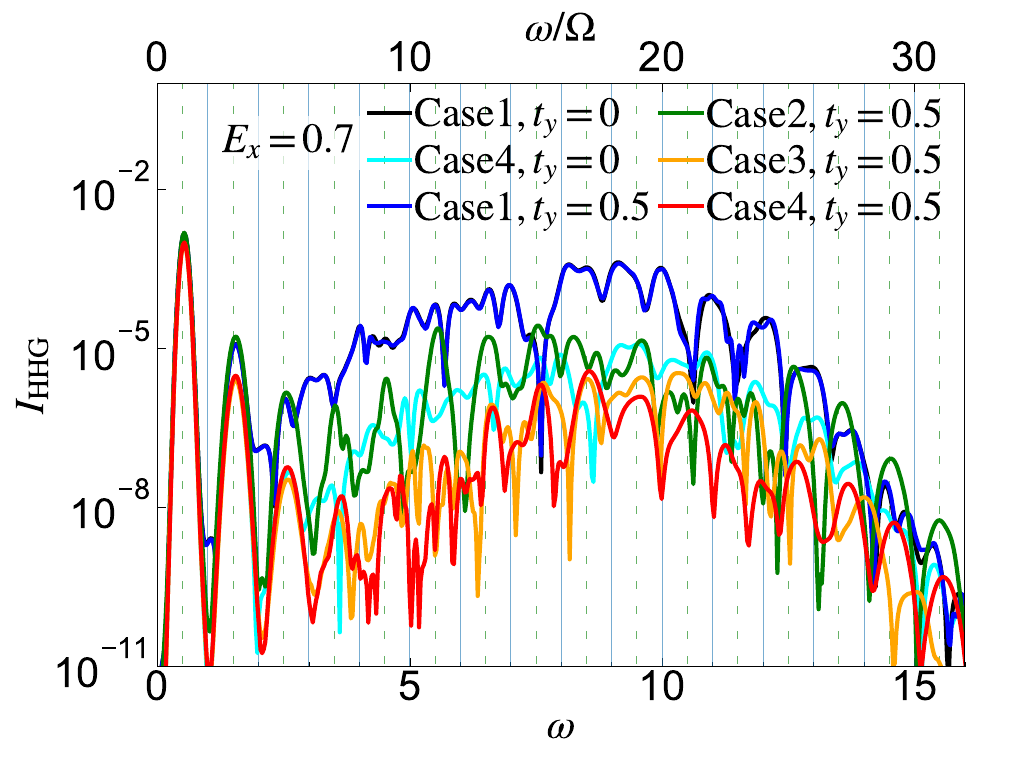} 
  \caption{HHG spectra of  the two-leg Mott insulator described by the effective model under the indicated conditions.
  We use $t_x=1$ and $U=8$, and set the excitation parameters to $\Omega=0.5$, $t_0=60$ and $\sigma_0=15$.
  The result for $t_y=0$ of Case 1) and that for $t_y=0.5$ of Case 1) almost overlap with each other.
}
  \label{fig:HHG_Heff}
\end{figure}

In Fig.~\ref{fig:HHG_Heff}, we show the HHG spectra evaluated from the effective models for different cases of $t_y=0$ and $t_y=0.5$. For $t_y=0$, peak structures are not prominent, similar to the HHG spectrum from the original Hubbard model. For $t_y=0.5$, in Case 4, where there is no parameter modification of the effective model, peak structures become clear and develop at $\omega= (2n+1)\Omega$ as expected. Thus, the effective model successfully describes the qualitative changes in the HHG features between $t_y=0$ and $t_y=0.5$.
To examine the effects of different physical processes, we compare the results of Cases 1) to 4) for $t_y=0.5$. See Fig.~\ref{fig:SW_model} for the details of each case. In Case 1), peak structures are not prominent and non-odd harmonic signals develop, indicating that the DH coherence remains long at this level. In Case 2), peak structures become prominent and odd-harmonic peaks develop, suggesting a reduction in pair coherence. In Case 3), peak structures become slightly more prominent compared to Case 1).

 \begin{figure}[t]
  \centering
    \hspace{-0.8cm}
    \vspace{0.0cm}
\includegraphics[width=94mm]{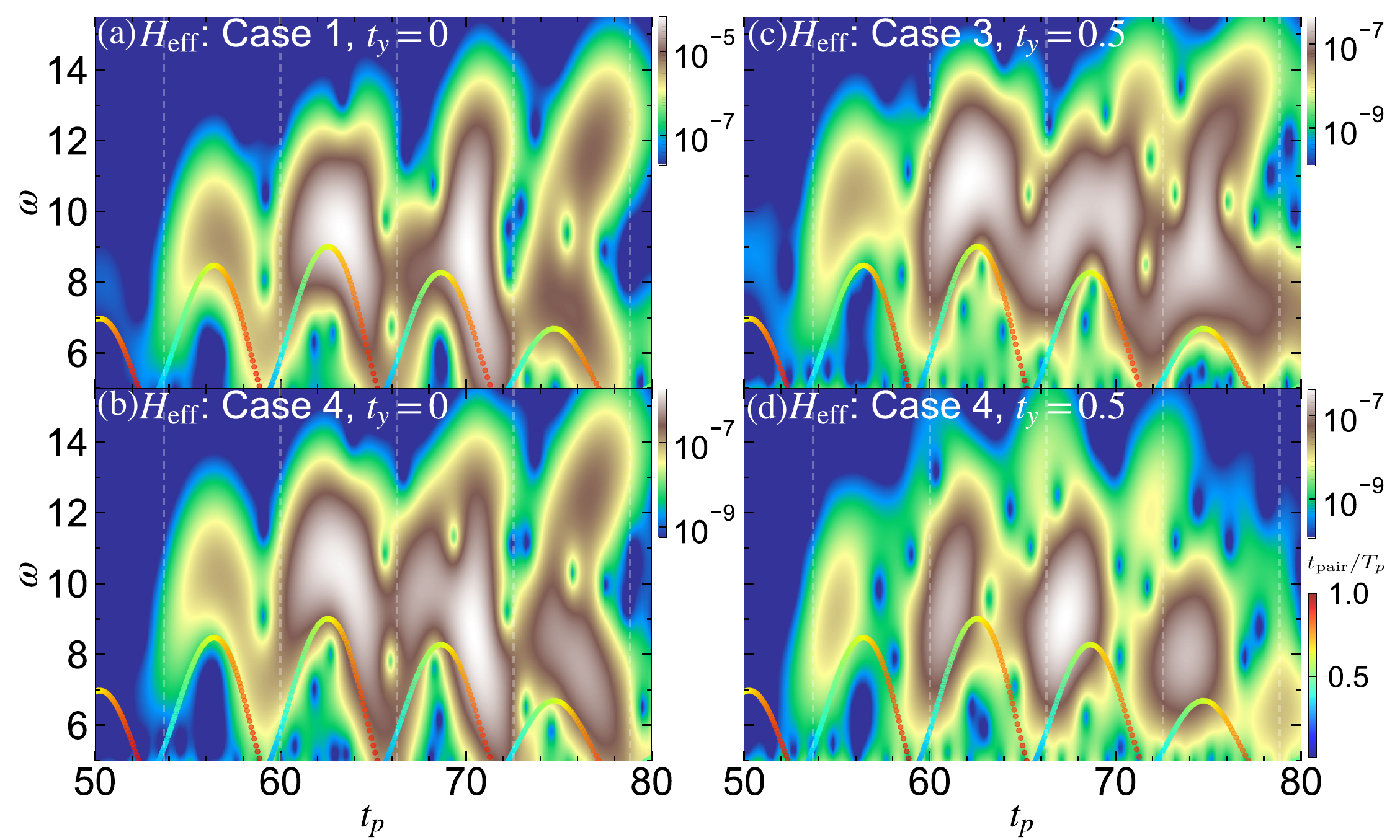} 
  \caption{Subcycle spectra of HHG in the two-leg Mott insulator described by the effective model.
  We use $t_x=1$ and $U=8$ and set the excitation parameters to $\Omega=0.5, E_x=0.7, t_0=60$ and $\sigma_0=15$.
  For the subcycle analysis, we use a Gaussian window with $\sigma_p=0.8$.
  (a) is for Case 1 at $t_y=0$, (b) is for Case 4 at $t_y=0$, (c) is for Case 3 at $t_y=0.5$ and (d) is for Case 4 at $t_y=0.5$
  The vertical dashed lines indicate the times when $A_x(t)=0$.
The multi-colored dots indicate the energy emitted at $t_p$ by the recombination of a doublon-holon pair with $\mathcal{E}_{\rm g}(k_x) = U -4t_x\cos(k_x)$.
The color indicates the time $t_{\rm pair}$ which elapsed between the creation and recombination of the pair ($T_p=2\pi/\Omega$).
}
  \label{fig:S_sub_Heff}
\end{figure}
To obtain detailed insights into different physical processes, we perform a subcycle analysis. In Figs.~\ref{fig:S_sub_Heff}(a,b), we show subcycle spectra for the effective models of Case 1) and Case 4) for $t_y=0$. In Case 1), the spectrum is well explained by the three-step model for a DH pair with $\mathcal{E}_{\rm g}(k_x) = U -4t_x\cos(k_x)$, while in Case 4, the subcycle spectrum is slightly off the prediction. The main origin is likely the shift of the energy band due to the spin-exchange term $\hat{H}_{\rm spin}$, which enhances the band gap. Still, in both cases, the results suggest that the main HHG contribution originates from the long-lived DH pairs.
In Figs.~\ref{fig:S_sub_Heff}(c,d), we show subcycle spectra for the effective models of Case 3 and Case 4 for $t_y=0.5$.
The corresponding results for Case 1 and Case 2 are shown in Fig.~3 in the main text.
In Case 3 and Case 4, the subcycle spectrum is slightly off the prediction from the three-step model for a DH pair with $\mathcal{E}_{\rm g}(k_x) = U -4t_x\cos(k_x)$,
compared to Case 1 and Case 2. Again the main origin is likely the shift of the energy band due to $\hH_{\rm spin}$.
Still, when we compare the results for Case 1 and Case 3, there is a shift of the signal to earlier time within one period of the field.
This suggests a reduction of the coherence of a DH pair due to the energetic coupling between spin and charge, i.e., the conversion of the kinetic energy of a doublon or a holon into spin exchange energy through the spin mismatch between the chains. 
This physics is directly related to the discussion in Ref.~\cite{Murakami2022PRL}, which considered a 1D single chain with a staggered magnetic field to emulate this type of coupling.
By comparing Case 2 and Case 3, we can see that the effect of the energetic coupling is less prominent compared to the non-energetic coupling, i.e. the dephasing due to the existence of the spin string, as pointed out in the main text.  Note that we can observe a prominent reduction of coherence in Case 3 for $t_y=1$ (not shown).
In Case 4, both effects are included and the coherence of the DH pair is reduced.

 \begin{figure}[t]
  \centering
    \hspace{-0.cm}
    \vspace{0.0cm}
\includegraphics[width=70mm]{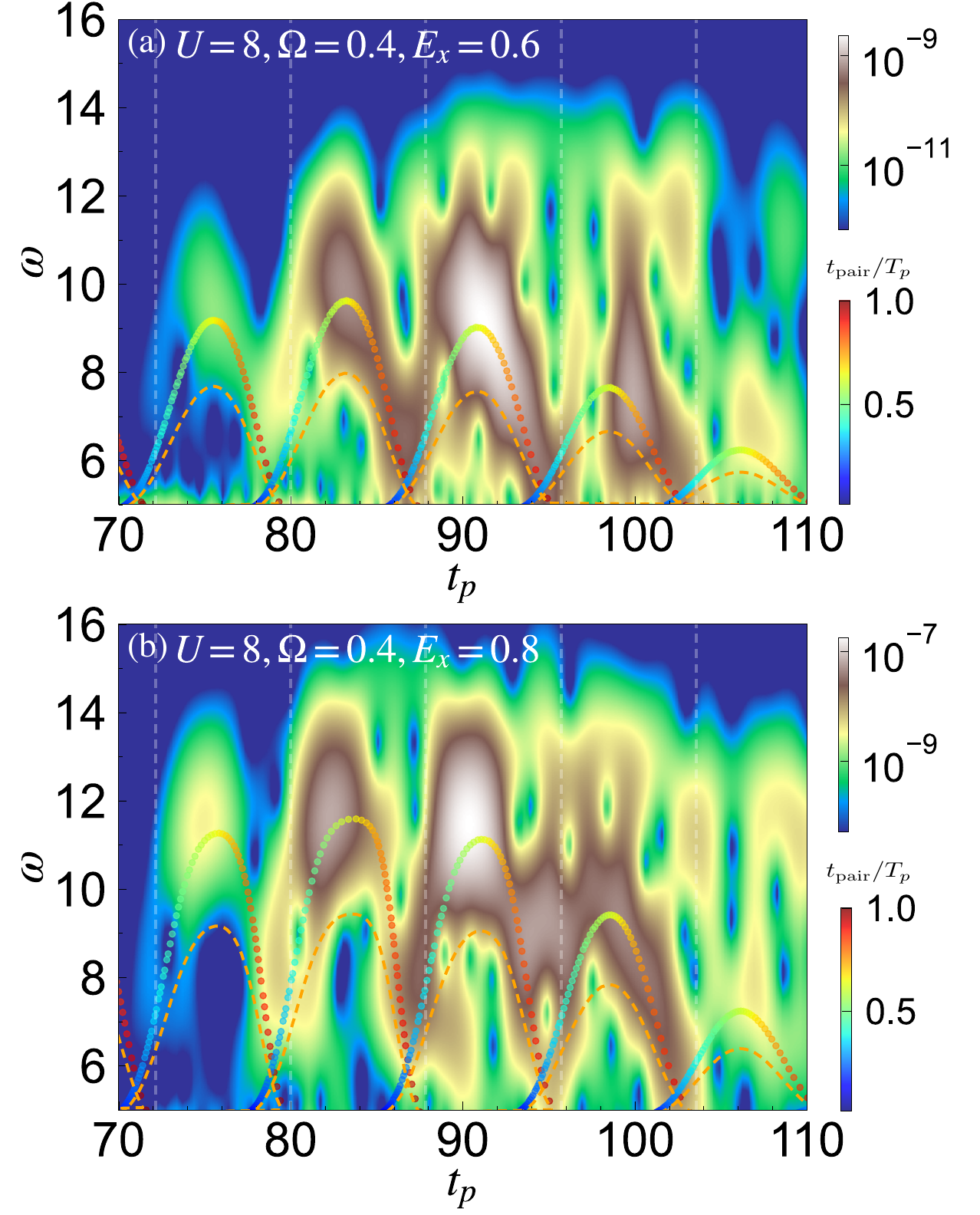} 
  \caption{ Subcycle spectra of HHG in the half-filled two-leg Hubbard model.
  We use $t_x=1,t_y=2$ and $U=8$ and set the excitation parameters to $\Omega=0.4,t_0=80$ and $\sigma_0=20$.
  For the subcycle analysis, we use a Gaussian window with $\sigma_p=0.8$.
  The vertical dashed lines indicate the times when $A_x(t)=0$.
  The multi-colored dots and the orange dashed lines indicate the prediction from the three-step model for Scenario 1 (hPolaron + ePolaron + magnon)
and for Scenario 2 (hPolaron + spin bag), respectively. 
We use the dispersion obtained from the strong-rung perturbation theory, see the text for details.
The dot color indicates the time $t_{\rm pair}$ which elapsed between the creation and recombination of the pair ($T_p=2\pi/\Omega$).
}
  \label{fig:subcycle_ty2}
\end{figure}

Now we provide further support for the HHG mechanism involving three elementary excitations in the strong-rung regime ($t_y=2$).
To emphasize the difference between Scenario 1 (hPolaron + ePolaron + magnon) and Scenario 2 (hPolaron + spin bag) within the three-step model, we reduce $\Omega$ to $\Omega=0.4$ (lower than in the main text).
In Fig.~\ref{fig:subcycle_ty2}, we show the corresponding subcycle spectra. One can see that the difference between the two scenarios is more pronounced and the intensity of the temporal radiation nicely follows the prediction from Scenario 1.

\subsection{Single-particle spectrum} \label{sec:Akw}

\begin{figure}[t]
     \centering
\includegraphics[width=85mm]{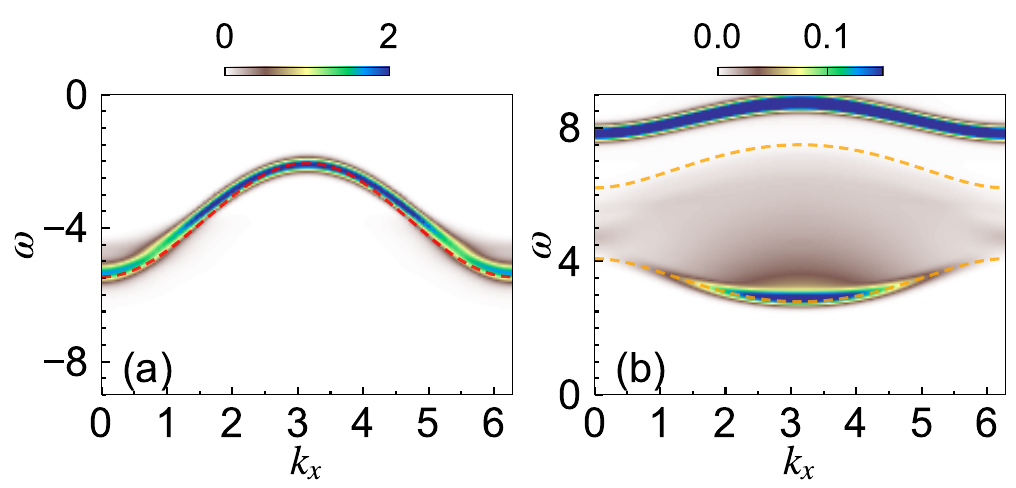}
\caption{Single-particle spectra $A(\bk,\omega)$ of the half-filled two-leg Hubbard model at $k_y=0$ evaluated with DMRG. 
Here we use  $U=8,t_x=1,t_y=2$ and $N_x=80$. 
The dashed line in (a) shows $-\mathcal{E}_{\rm hP}(k_x,k_y=0) +\Delta \omega$ with $\Delta\omega = -0.1$. 
The dashed lines in (b) show the lower and upper bounds of the continuum (Eq.~\eqref{eq:contunum}) for the magnon ($\mathcal{E}_{\rm M}(k_x,k_y=\pi)$) and the ePolaron ($\mathcal{E}_{\rm eP}(k_x,k_y=\pi)$) shifted by $\Delta\omega = 0.3$. } 
\label{fig:Akw_ty2}
\end{figure}

In order to obtain information on the relevant elementary excitations, we calculate the single-particle spectrum, which is
defined as $A(\bk,\omega) = -\frac{1}{\pi} {\rm Im} G^R(\bk,\omega)$.
Here $G^R(\bk,\omega)$ is the Fourier transform of the space and time-dependent retarded Green's function $G^R_{\bm{ij}\sigma}(t) = -i\theta(t)\langle [\hc_{\bm i\sigma}(t),\hc^\dagger_{\bm j\sigma}(0)]_+\rangle$. 
We evaluate $G^R_{\bm{ij}\sigma}(t)$ for a large but finite-size system using DMRG as mentioned in Sec.~\ref{sec:DMRG}.
For the Fourier transform of the time variable, we apply the envelope function $F_G(t,\sigma) = \exp(-\frac{t^2}{2\sigma^2})$.
The single-particle spectrum of the two-leg Hubbard model has been systematically studied using a similar DMRG technique in Ref.~\cite{Feiguin2019PRB}.

Some results are shown in Fig.~4 of the main text, where we use $\sigma=8$. 
In Fig.~\ref{fig:Akw_ty2}, we show a magnified view of $A(\bk,\omega)$ for  $U=8$ and $t_y=2$ at $k_y=0$, to check the consistency and reveal the correction to the strong-rung perturbation analysis (Sec.~\ref{sec:rung_perturb}).
In these figures,  we show results obtained by calculating the time evolution with the second-order Trotter-Suzuki decomposition, a time-step $dt=0.02$, and cut-off dimension $D=1000$.
For $\omega\leq0$ (Fig.~\ref{fig:Akw_ty2}(a)), we see the signal from a hPolaron, which is well explained by $-\mathcal{E}_{\rm hP}(k_x,k_y=0)$ (Eq.~\eqref{eq:hole_polaron}) with a small shift $\Delta \omega= -0.1$.
This shift can be attributed to higher-order corrections.
For $\omega\geq0$ (Fig.~\ref{fig:Akw_ty2}(b)), we see the continuum consisting of a magnon and an ePolaron. 
The boundary of the continuum is well explained by Eq.~\eqref{eq:contunum} for $\mathcal{E}_{\rm M}(k_x,k_y=\pi)$ (Eq.~\eqref{eq:magnon}) and $\mathcal{E}_{\rm eP}(k_x,k_y=\pi)$ (Eq.~\eqref{eq:electron_polaron})
with a small shift $\Delta \omega= -0.3$, which can also be attributed to the higher-order corrections.
Note that both the magnon and the e-Polaron have $k_y=\pi$, so that the combination of them leads to $k_y=0$.
The intensity of the spectrum at the bottom of the magnon-polaron continuum is especially strong. 
In Ref.~\cite{Feiguin2019PRB}, this feature has been attributed to a loosely bound state (a new composite state) of a magnon-polaron pair, called a spin-bag.
These results suggest that the perturbation theory captures well the kinetic properties of the elementary excitations at $t_y=2$.

\subsection{Results for semiconductors}
\begin{figure}[t]
     \centering
\includegraphics[width=70mm]{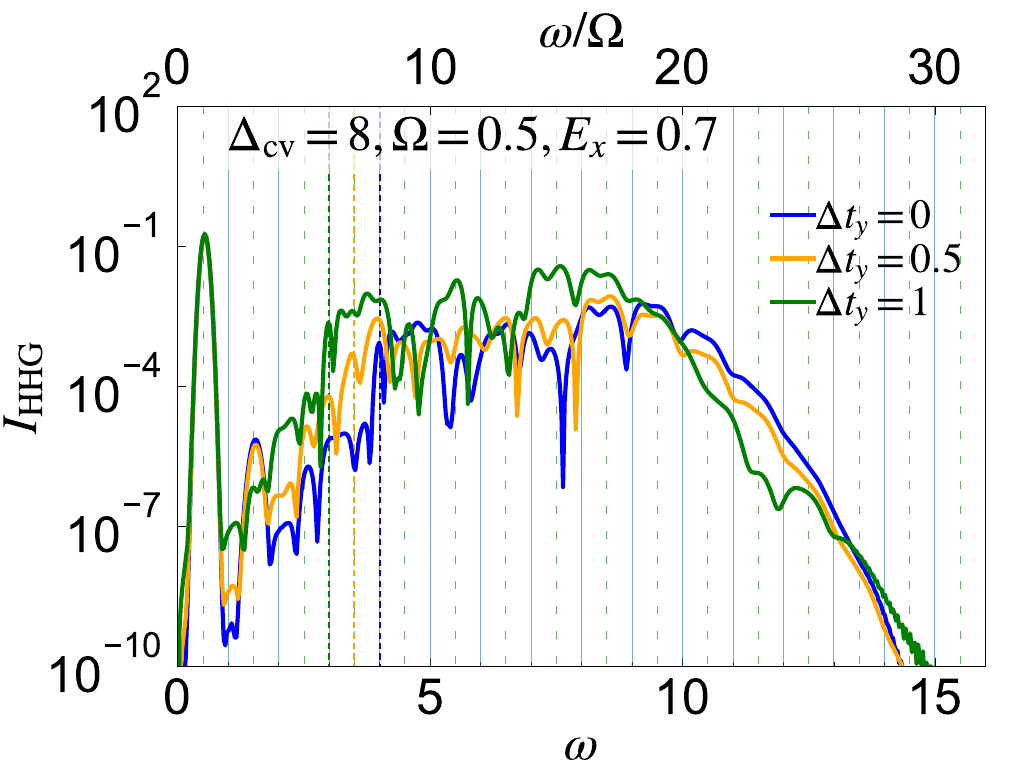}
\caption{HHG spectra for the semiconductor ~\eqref{eq:Ham_semicon} evaluated using the SBE. 
We use $\Delta_{\rm cv}=8$, $t_{{\rm c},x}=1$, $t_{{\rm v},x}=-1$ and $D_x=0.5$, and choose the excitation parameters as $E_x=0.7,\Omega=0.5, t_0=60$ and $\sigma_0=15$. 
The vertical solid (dotted) lines indicate $\omega = 2n \Omega$ ($\omega = (2n+1)\Omega$) with $n\in\mathbb{N}$.
The colored vertical dashed lines indicate the band gap for the corresponding hoppings.} 
\label{fig:HHG_semicon}
\end{figure}

\begin{figure}[t]
     \centering
\includegraphics[width=70mm]{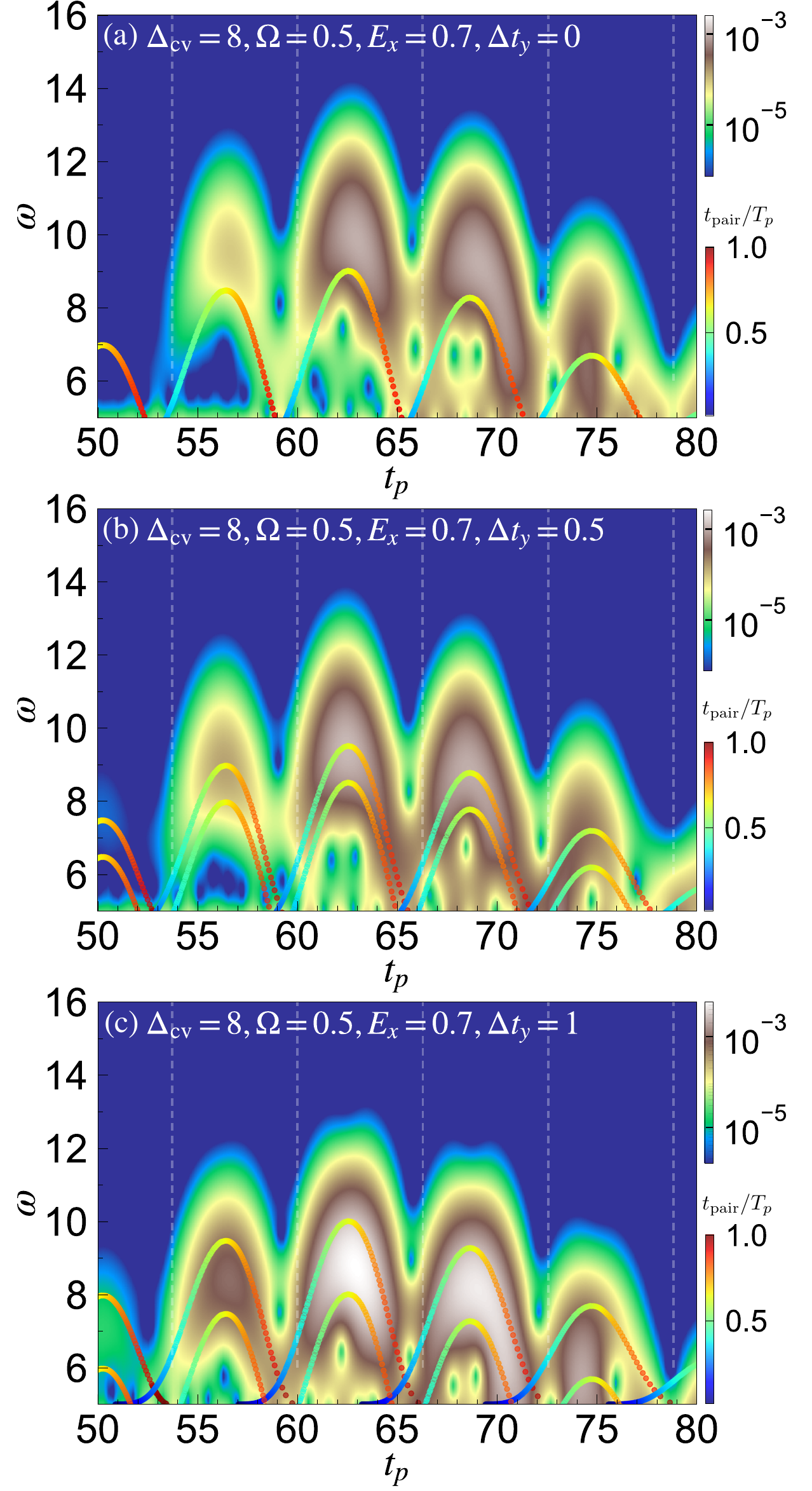}
\caption{Subcycle spectra for the semiconductor \eqref{eq:Ham_semicon} and the indicated parameters. 
We use $\Delta_{\rm cv}=8$, $t_{{\rm c},x}=1$, $t_{{\rm v},x}=-1$ and $D_x=0.5$, and choose the excitation parameters as $E_x=0.7,\Omega=0.5, t_0=60$ and $\sigma_0=15$. 
For the subcycle analysis, we use a Gaussian window with $\sigma_p=0.8$.
The vertical dashed lines indicate the times where $A_x(t)=0$.
The rainbow-colored dots indicate the energy emitted at $t_p$ by the recombination of an electron-hole pair, as obtained from the three-step model.
The dot color indicates the time $t_{\rm pair}$ which elapsed between the creation and recombination of the pair  in units of $T_p=2\pi/\Omega$.
} 
\label{fig:semicon_subcycle}
\end{figure}

In order to reveal the differences to semiconductors, we introduce a standard semiconductor model~\cite{Vampa2015PRB,Murakami2021PRB} on the two-leg ladder.
Here we assume that the Wannier orbitals for the conduction band and the valence band are 
localized 
on the same site.
In the dipolar gauge~\cite{Li2020,Michael2021PRB}, the semiconductor Hamiltonian reads
\eqq{
\hH(t) &= \sum_{{\bm i},a} \Delta_a \hn_{{\bm i}a}  - \sum_{i_x,a}t_{a,y} [\hc^\dagger_{i_x0a} \hc_{i_x1 a} + h.c. ]   \nonumber \\
& -\sum_{{\bm i},a}t_{a,x}  [e^{iqA_x(t)}\hc^\dagger_{{\bm i}+{\bm e}_xa} \hc_{{\bm i}a} + h.c. ]   \nonumber \\
& -E_x(t) D_x\sum_{{\bm i},a} \hc^\dagger_{{\bm i}a}\hc_{{\bm i}\bar{a}}. \label{eq:Ham_semicon}
}
 $a$ is the index for the conduction band (c) and the valence band (v), $\bar{a}$ indicates the band different from $a$,
 $\hc^\dagger_a$ is the creation operator of an electron on band $a$ and $ \hn_{{\bm i}a} = \hc^\dagger_{{\bm i}a}\hc_{{\bm i} a}$.
 $t_{a,\alpha}$ with $\alpha=x,y$ is the hopping parameter of the $a$-band electron along the $\alpha$-direction.
 $\Delta_a$ is the energy level of the $a$-band.
 $D_x$ represents the interband dipole moment for the $x$ direction, which we assume to be local~\cite{Vampa2015PRB}.
The corresponding interband polarization operator is defined as $\hP_{x,{\rm er}} \equiv  D_x\sum_{{\bm i},a} \hc^\dagger_{{\bm i}a}\hc_{{\bm i}\bar{a}}$. 
Here, we only consider the electric field along the $x$ direction, and the light-matter coupling is taken into account by the Peierls phase and the interband dipole term. For simplicity, we ignore the spin degree of freedom.

In momentum space, we have
\eqq{
\hH(t) &= \sum_{{\bm k},a} [\mathcal{E}_a(k_x-qA_x(t),k_y)+\Delta_a] \hc^\dagger_{\bm k} \hc_{\bm k}  \nonumber \\
&  -E_x(t) D_x \sum_{{\bm k},a}\hc^\dagger_{{\bm k},a} \hc_{{\bm k},{\bar a}} ,
}
where $\hc^\dagger_{{\bm k},a}=\frac{1}{\sqrt{N}} \sum_{{\bf i}} e^{i {\bm k}\cdot{\bm i}} \hc^\dagger_{{\bm i},a}$, ${\bm k}=(k_x,k_y)$ and ${\bm i}=(i_x,i_y)$.  Note that $k_y=0$ or $k_y=\pi$. The band dispersion becomes
\eqq{
\mathcal{E}_a(k_x,k_y) &= -2t_{a,x} \cos(k_x) -t_{a,y} \cos(k_y). 
}
Note that the shape of the band dispersion along $k_x$ is not modified by the hopping along $y$, and that it is the same for $k_y=0$ and $k_y=\pi$.
The only difference is the energy level, controlled by $\Delta t_{y} \equiv |t_{{\rm c},y}-t_{{\rm v},y}|$. Thus, the hopping along $y$ hardly affects the kinematics of the electron-hole pair in the HHG process.
This situation is very different from the Mott insulator on the two-leg ladder, where hopping along $y$ yields drastic changes in the excitation structure,  
such as the formation of polarons and spin bags.

To demonstrate the effects of hopping along $y$ in semiconductors, we simulate the time evolution using the semiconductor Bloch equation (SBE)~\cite{Vampa2015PRB}
assuming that the valence band is initially fully occupied.
Here, we set $\Delta_{\rm c}-\Delta_{\rm v}\equiv \Delta_{\rm cv}=8$, $t_{{\rm c},x}=1$ and $t_{{\rm v},x}=-1$, to make the energy range of the bands similar to that of the Mott insulator in the main text.
We use the same shape of the electric field pulse as in the main text and set the dipole moment $D_x$ to 0.5.
The HHG intensity is evaluated as $I_{\rm HHG}(\omega)=|\omega J_{x,\rm ra}(\omega) + \omega^2 P_{x,\rm er}(\omega)|^2$, where  $J_{x,\rm ra}$ is the intraband current and $P_{x,\rm er}$ is the interband polarization.

The resultant HHG spectrum is shown in Fig.~\ref{fig:HHG_semicon}.
Here we do not add phenomenological dephasing terms in the SBE.
The shape of the HHG spectrum is hardly affected by hopping along the $y$ direction.
In addition, the intensity is increased for nonzero $\Delta t_{{\rm cv},y}$, compared to the case with $\Delta t_{{\rm cv},y}=0$, due to the reduction of the band gap.
We also show the corresponding subcycle analysis in Fig.~\ref{fig:semicon_subcycle}.
Here we only focus on the contribution from the interband current, i.e. $P_{x,\rm er}$, which provides the dominant contribution to HHG above the gap.
The profile of the subcycle spectrum is hardly affected, except for the shift of the signal due to the shift of the bands.
This result demonstrates the absence of a change in the coherence of the electron-hole pairs.
All of these features are in stark contrast to the findings for the Mott insulator.

\bibliography{HHG_Ref}

\end{document}